\documentclass[aps, prb,8pt, citeautoscript, superscriptaddress, twocolumn, amsmath]{revtex4-2}

\usepackage[T1]{fontenc} 
\usepackage{graphicx}
\usepackage{natbib}
\usepackage{siunitx}
\usepackage{bm} 
\usepackage{enumitem}
\usepackage{textcomp}
\setlist{nosep} 
\usepackage[colorlinks=true,citecolor=blue,linkcolor=blue]{hyperref}
\usepackage[utf8]{inputenc} 
\usepackage[dvipsnames]{xcolor} 

\setcitestyle{super}

\usepackage{color}

\newcommand{\ket}[1]{\ensuremath{\left| #1 \right\rangle}}

\newcommand{\abs}[1]{\ensuremath{\left| #1 \right|}}

\newcommand{\Sone}{\ensuremath{S1_{\rm Air}\,}} 
\newcommand{\Stwo}{\ensuremath{S2_{\rm Encap}\,}}	
\newcommand{\Sthree}{\ensuremath{S3_{\rm Air}\,}}	
\newcommand{\Sfour}{\ensuremath{S4_{\rm Air}\,}} 

\begin{document}

\title{Nanoscale magnetism and magnetic phase transitions in atomically thin CrSBr}

\author{Märta A. Tschudin}
\thanks{These authors contributed equally}
\affiliation{Department of Physics, University of Basel,  Basel, Switzerland }

\author{David A. Broadway}
\email{david.broadway@rmit.edu.au}
\thanks{These authors contributed equally}
\affiliation{Department of Physics, University of Basel,  Basel, Switzerland }

\author{Patrick Reiser}
\affiliation{Department of Physics, University of Basel,  Basel, Switzerland }

\author{Carolin Schrader}
\affiliation{Department of Physics, University of Basel,  Basel, Switzerland }

\author{Evan J. Telford}
\affiliation{Department of Physics, Columbia University, New York, NY, USA}
\affiliation{Department of Chemistry, Columbia University, New York, NY, USA}

\author{Boris Gross}
\affiliation{Department of Physics, University of Basel,  Basel, Switzerland }

\author{Jordan Cox}
\affiliation{Department of Chemistry, Columbia University, New York, NY, USA}

\author{Adrien E. E. Dubois}
\affiliation{Department of Physics, University of Basel,  Basel, Switzerland }
\affiliation{QNAMI AG, Hofackerstrasse 40 B, Muttenz CH-4132, Switzerland}

\author{Daniel G. Chica}
\affiliation{Department of Chemistry, Columbia University, New York, NY, USA}
 
 \author{Ricardo Rama-Eiroa}
 \affiliation{Donostia International Physics Center (DIPC), 20018 Donostia-San Sebastián, Basque Country, Spain\looseness=-1}
\affiliation{Institute for Condensed Matter Physics and Complex Systems, School of Physics and Astronomy, The University of Edinburgh, Edinburgh, EH9 3FD, United Kingdom}

\author{Elton J. G. Santos}
\affiliation{Institute for Condensed Matter Physics and Complex Systems, School of Physics and Astronomy, The University of Edinburgh, Edinburgh, EH9 3FD, United Kingdom}
\affiliation{Higgs Centre for Theoretical Physics, The University of Edinburgh, Edinburgh EH9 3FD, United Kingdom}
\affiliation{Donostia International Physics Center (DIPC), 20018 Donostia-San Sebastián, Basque Country, Spain\looseness=-1}

\author{Martino Poggio}
\affiliation{Department of Physics, University of Basel,  Basel, Switzerland }
\affiliation{Swiss Nanoscience Institute, University of Basel, Basel, Swizterland}

\author{Michael E. Ziebel}
\affiliation{Department of Chemistry, Columbia University, New York, NY, USA}

\author{Cory R. Dean}
\affiliation{Department of Physics, Columbia University, New York, NY, USA}

\author{Xavier Roy}
\affiliation{Department of Chemistry, Columbia University, New York, NY, USA}

\author{Patrick Maletinsky}
\email[]{patrick.maletinksy@unibas.ch}
\affiliation{Department of Physics, University of Basel,  Basel, Switzerland }

\date{\today}

\maketitle

\textbf{
Since their first observation in 2017\,\cite{Gong2017, Huang2017}, atomically thin van der Waals  (vdW) magnets have attracted significant fundamental, and application-driven attention\,\cite{Gibertini2019, Gong2019, Wang2022}. 
However, their low ordering temperatures, $T_c$\,\cite{Gong2017, Huang2017,Thiel2019, Broadway2020}, sensitivity to atmospheric conditions\,\cite{Liu2019, Shcherbakov2018} and difficulties in preparing clean large-area samples\,\cite{Cao2015} still present major limitations to further progress.
The remarkably stable\,\cite{Telford2022}, high-$T_c$ vdW magnet CrSBr\,\cite{Telford2020,Wang2020} has the potential to overcome these key shortcomings, but its nanoscale properties and rich magnetic phase diagram remain poorly understood. 
Here we use single spin magnetometry\,\cite{Maletinsky2012} to 
quantitatively characterise saturation magnetization, magnetic anisotropy constants, and magnetic phase transitions in few-layer CrSBr by direct magnetic imaging. 
We show pristine magnetic phases, devoid of defects on micron length-scales, and demonstrate remarkable air-stability down the monolayer limit.
We address the spin-flip transition in bilayer CrSBr by direct imaging of the emerging antiferromagnetic (AFM) to ferromagnetic (FM) phase wall and elucidate the magnetic properties of CrSBr around its ordering temperature. 
Our work will enable the engineering of exotic electronic and magnetic phases in CrSBr and the realisation of novel nanomagnetic devices based on this highly promising vdW magnet.
}

\begin{figure*}[ht]
	\centering
	\includegraphics{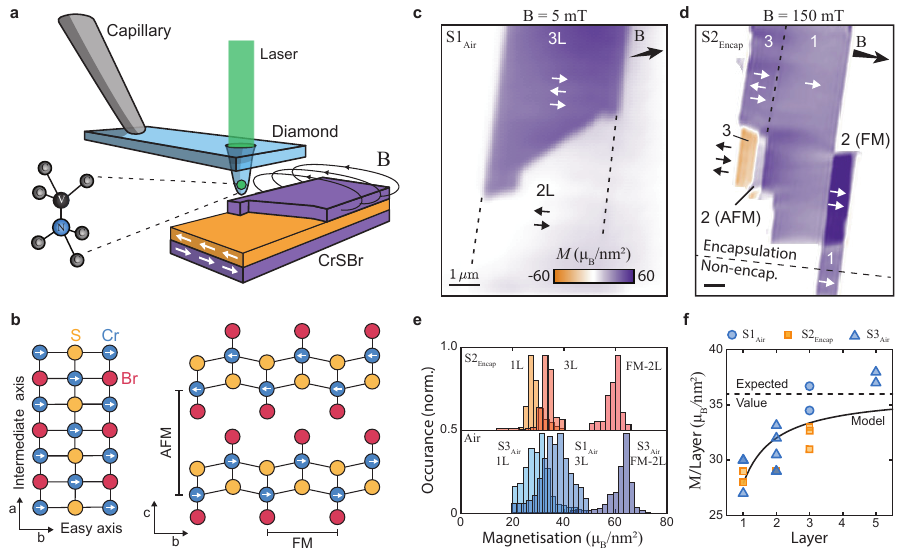}
	\caption{\textbf{Magnetic characterization of few-layer CrSBr}. 
	\textbf{a} Illustration of the experiment involving an all-diamond tip containing a single-spin magnetometer that is scanned over the sample to spatially image magnetic stray fields.
    \textbf{b} Crystallographic structure of CrSBr, where the magnetic easy (intermediate) axis is aligned along the $b$($a$)-axis of the crystal respectively.
	\textbf{c} Magnetization image of a non-encapsulated bilayer and trilayer flake (Sample 1, \Sone), obtained in a low bias magnetic field $|B_{\rm ext}|=5~$mT.
	\textbf{d} Magnetization image of an encapsulated multi-layer flake (Sample 2, \Stwo), obtained in a bias magnetic field $|B_{\rm ext}|=150~$mT strong enough to induce FM ordering in some bilayer sections. 
    \textbf{e} Exemplary magnetization histograms of encapsulated (top panel) and non-encapsulated (bottom panel) flakes of different thicknesses taken from additional datasets\,\cite{SOM}. 
    \textbf{f} Extracted magnetization per CrSBr layer for encapsulated (orange) and non-encapsulated (blue) flakes as a function of the number of layers. The dashed line is the expected value ($36~\mu_B/$nm$^2$) from bulk measurements and the solid black line is a simplified model fit (see text).   
 }
\label{Fig:intro}
\end{figure*}

Heterostructures based on 2D vdW materials have had a profound impact on our understanding and control of a vast range of electronic and optical phenomena in condensed matter physics\,\cite{Cao2018a,Xu2018a,Shimazaki2021a}.
By comparison, explorations of 2D vdW magnets, which are typically highly fragile and exhibit low ordering temperatures, are still in their infancy, despite remarkable progress in observing exotic forms of magnetism, including 2D-XY magnetism\,\cite{Bedoya2021}, orbital ferromagnetism\,\cite{Tschirhart2021}, and Moir\'e magnetism\,\cite{Song2021}. 
The magnetically ordered vdW semiconductor CrSBr has emerged as a highly attractive candidate to overcome these shortcomings. 
Next to its in-plane, A-type AFM ordering (Fig.\,\ref{Fig:intro}\,\textbf{a}), CrSBr exhibits unusual transport properties\,\cite{Wu2022a}, an intriguing interplay between magnetic and optical properties\,\cite{Dirnberger2023a}, and a remarkable tunability of its magnetism by strain\,\cite{Cenker2022}. 
Importantly, and unlike other 2D magnetic vdW materials, CrSBr shows remarkable structural stability\,\cite{Telford2022}, magnetically orders at a relatively high (N\'eel) temperature $T_N\approx132~$K, with evidence for FM intralayer interactions persisting to even higher temperatures\,\cite{Telford2020,Lee2021,Telford2022}.  
Yet, little is thus far known about the nanoscale properties of CrSBr and how its various magnetic phases develop and transition in few layer thin samples, which have so far been addressed only by non-quantitative\,\cite{Lee2021, Moro2022} and invasive\,\cite{Rizzo2022} imaging methods.

Here, we employ single-spin scanning magnetometry\,\cite{Maletinsky2012} using an individual nitrogen vacancy (NV) center in diamond\,\cite{Balasubramanian2009a} -- a nanoscale imaging technique that is non-invasive and 
sensitive enough to image magnetism in vdW monolayers\,\cite{Thiel2019} -- to explore magnetic order in CrSBr in a quantitative way.  
The NV spin is situated at the tip of an atomic force microscope and scanned across the CrSBr flakes (Fig.\,\ref{Fig:intro}\,\textbf{a}) to quantitatively image magnetic stray fields through Zeeman-shifts of the spin's energy levels.
The NV thereby enables a quantitative\,\cite{Rondin2014a} determination of $B_{\rm NV}$ -- the projection of the total magnetic field onto the NV axis -- which in turn allows us to determine CrSBr's sample magnetization and magnetic anisotropy energy, down to the monolayer limit.
Furthermore, building on the high stability and spatial resolution of our approach, we provide direct visualizations of key magnetic phase transitions in mono- and bilayers of CrSBr.

\begin{figure*}[ht]
	\centering
	\includegraphics{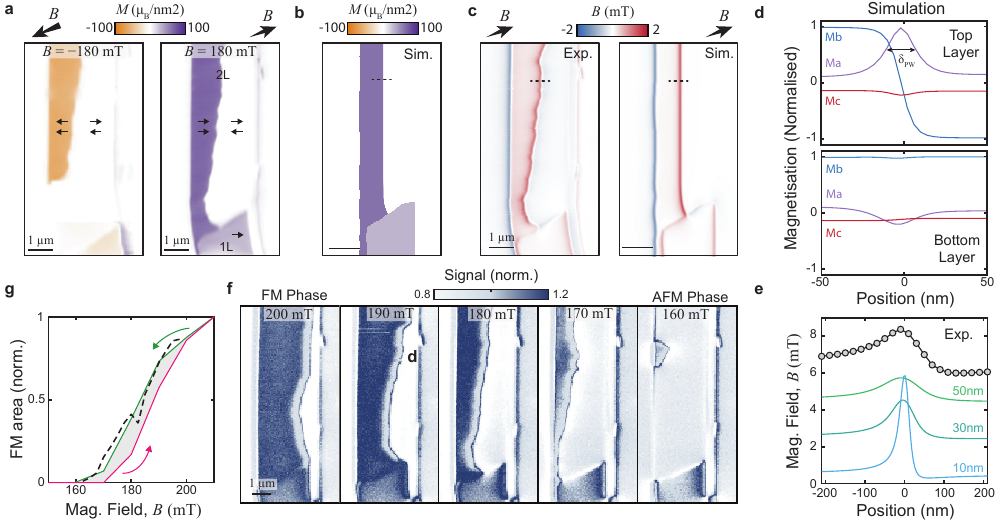}
 \caption{\textbf{Imaging the FM to AFM spin-flip transition in bilayer CrSBr.} 
 \textbf{a} Image of coexisting FM and AFM phases in a CrSBr bilayer obtained at $|B_{\rm ext}|=\pm 180~$mT after demagnetizing the sample from higher fields, with a magnetic field axis oriented along $(\theta, \phi) = (53^\circ, 16^\circ)$ (see text).
 The images were obtained by magnetization reconstruction from raw $|B_{\rm NV}|$ data\,\cite{SOM}. 
 \textbf{b} Micromagnetic simulation of the phase wall under the same conditions as the right panel in \textbf{a}. 
 \textbf{c} Experimental magnetic field image from the measurement in \textbf{a} and simulated magnetic field from \textbf{b}.
 \textbf{d} Cross sections of the magnetization vector across the PW  (dashed line in \textbf{b}) for both the top and bottom layers.  
 \textbf{e} Comparison of the experimental measured magnetic field line cut across the PW  (dashed lines in \textbf{c}) and the simulation with various NV-sample standoff distances, offset for clarity.  
\textbf{f} Qualitative NV magnetometry images\,\cite{SOM} of the movement of the FM-AFM phase wall when demagnetising the sample from the FM to the AFM state.  
 \textbf{g} Magnetic hysteresis of the bilayer sample, calculated as a ratio of the total area in the FM state versus the AFM state\,\cite{SOM}. 
 The dashed line corresponds to data from an additional measurement run\,\cite{SOM}.
 }
 \label{Fig: Phase Transition}
\end{figure*}

We start by investigating few-layer CrSBr samples at $T\approx4~$K and at low magnetic fields ($B\approx 5$~mT), where the CrSBr layers exhibit AFM interlayer alignment\,\cite{Telford2020}, 
and where a nonzero net sample magnetization can thus only be expected from odd numbers of layers.  
Our samples were fabricated by mechanically exfoliating bulk CrSBr crystals onto a Si/SiO$_2$ substrate, where a subset of the resulting flakes were encapsulated in hexagonal boron nitride (hBN) and the rest remained bare, non-encapsulated flakes.
CrSBr typically exfoliates into large, near-rectangular flakes, where the long (short) edges of the rectangle correspond to the $a$ ($b$) crystallographic axes (Fig.\,\ref{Fig:intro}\,\textbf{b}).
Axis $b$ corresponds to the magnetic easy-axis, where magnetization alignment along $a$ is suppressed by an energy penalty that has not been determined for few-layer samples thus far. 

We perform scanning NV magnetometry across our CrSBr flakes to obtain magnetic images\,\cite{SOM}, from which we reconstruct the underlying magnetization structure using a neural network approach\,\cite{Dubois2022}. Figures\,\ref{Fig:intro}\,\textbf{c},\,\textbf{d} show representative data on flakes containing mono-, bi-, and tri-layer sections of CrSBr.  
Throughout, these data reveal largely uniform magnetizations, which are well-aligned with the magnetic easy axis $b$.

We determine the average magnetization of CrSBr flakes through histograms of the reconstructed magnetization maps (Fig.\,\ref{Fig:intro}\,\textbf{e}).
Applying this approach to a series of CrSBr flakes with increasing layer numbers, we observe an initial increase in the magnetization per layer, until reaching the expected saturation magnetization at $N\gtrsim5$ layers to the expected bulk value of $36~\mu_B/$nm$^2$ (Fig.\,\ref{Fig:intro}\,\textbf{f})\,\cite{SOM}, where $\mu_B$ is the Bohr magneton. 
To validate our findings, we independently determined CrSBr sample magnetizations by performing analytic fits to the measured magnetic field emerging from flake edges, which agrees with our initial approach\,\cite{SOM}.
The observed increase of magnetization with the number of layers suggests that interlayer exchange coupling\,\cite{bo2023calculated} plays a relevant role in stabilizing intralayer FM ordering.
Indeed, the scaling of layer magnetization, $M_l$, with $N$ fits well to an empirical model (black line in Fig.\,\ref{Fig:intro}\,\textbf{f}) that assumes a magnetization reduction, $\epsilon$, for the outermost layer of Cr atoms, but constant magnetization, $M_0$, for the remaining Cr layers $M_l = M_0(N - 2\epsilon)/N$.
The model yields a bulk magnetization $M_0 = 36.1(1)~\mu_B/$nm$^2$, and a reduction factor of $11(4)\%$ for the outermost layers.
We note that our findings are reproduced and near-identical whether we investigate encapsulated or non-encapsulated flakes (Fig.\,\ref{Fig:intro}\,\textbf{f}).
This attests to the remarkable air-stability of magnetism in CrSBr, which we observe down to monolayers that had been left exposed to ambient conditions for days. 

In the presence of moderate in-plane magnetic fields, CrSBr undergoes a metamagnetic transition from inter-layer AFM to FM configuration.
For fields applied along the $b$-axis, this  corresponds to a sharp spin-flip transition, while for fields along the $a$-axis, a broad transition range indicates continuous canting of the spins\,\cite{Cenker2022,Telford2022}.
Past observations of non-symmetric hysteresis curves around the spin-flip transition\,\cite{Ye2022}, suggest the possibility that FM and AFM phases can coexist during the transition. 
However, whether such a mixed phase indeed exists and how the spin-flip transitions occur on the microscopic level in CrSBr remains unknown.

To address the physics of this spin-flip transition and the possibility of coexisting AFM/FM phases, we initialise a CrSBr bilayer into the FM configuration by applying a magnetic field $B_{\rm ext}=230~$mT along the NV axis, such that the in-plane projection of $B_{\rm ext}$ is approximately aligned with the sample's $b$-axis.
The field's polar and azimuthal angles amount to $(\theta, \phi) = (53^\circ, 16^\circ)$, where $\theta=0^\circ$ corresponds to the sample normal and $\phi=0^\circ$ to the horizontal axis in all images.
We subsequently decrease $B_{\rm ext}$ in $10~$mT steps and perform magnetic imaging to identify the AFM to FM flipping field.
We observe that at $B_{\rm ext}\approx200~$mT, an AFM ordered region develops in the flake. 
Fig.\,\ref{Fig: Phase Transition}\,\textbf{a} shows a full magnetization map obtained at $B_{\rm ext}\approx 180~$mT that evidences the coexistence of FM and AFM phases during the spin-flip transition.
This observation is reproducible and analogously occurs at inverted magnetic fields. 
The boundary between FM and AFM phases that we term a ``phase wall'' (PW) is stable over several days, which is consistent with the spin-flip transition exhibiting a large imaginary AC magnetic susceptibility that implies irreversible domain wall movement\,\cite{SOM}.

To provide a deeper understanding of the mechanism behind the formation of the AFM-FM PW, we performed micromagnetic simulations, taking into account the known magnetic properties of CrSBr\,\cite{Vansteenkiste2014, Leliaert2017, Exl2014} (details in Ref.\,\cite{SOM}). 
For a homogeneous CrSBr bilayer, we are unable to stabilize a PW in our simulations, i.e. the flake 
immediately switches from FM to AFM (or vice versa) within one field step at \SI{160}{\milli\tesla} (\SI{195}{\milli\tesla}).
We conjecture that inhomogeneities in the sample contribute to the stabilisation of the PW.
We model these by imposing that the interlayer exchange coupling strength $A_{ex,z}$ exhibits a linear variation 
 along the $b$-axis that amounts to $24~$\% across the flake width of $3.3~\mu$m. 
While other inhomogeneities or pinning centers could also explain the stability of the PW, a variation of $A_{ex,z}$ appears plausible, as it can be induced by strain-gradients\,\cite{Cenker2022} that commonly occur in van der Waals structures\,\cite{Jain2018a}.
Using literature values\,\cite{Cenker2022}, we estimate that the variation in $A_{ex,z}$ we impose can be induced by a realistic $0.4\%$ strain variation across the CrSBr flake.
Figures\,\ref{Fig: Phase Transition}\,\textbf{b} and \textbf{c} show the resulting simulated sample magnetization and stray field pattern in the presence of the  PW, where the latter shows good agreement with our experimental data.

Our simulations further show that the magnetization rotation across the PW is largely confined to the $a$-$b$ plane, with a sense of rotation that is set by the nonzero projection of $B_{\rm ext}$ onto the $a$-axis.
The simulated PW extends over a  (Bloch) width of $\delta_{PW} = 18~$nm (Fig.\,\ref{Fig: Phase Transition}\,\textbf{d}), which is below the spatial resolution of our method. 
Yet, the experimental data show good agreement with the model and their comparison allows us to estimate the NV-sample distance (that sets the spatial resolution) to $\sim50~$nm (Fig.\,\ref{Fig: Phase Transition}\,\textbf{e}).

To further investigate the spin-flip transition in bilayer CrSBr, we follow the movement of the PW through the flake as the spin-flip transition occurs. 
For this, we use a qualitative ``dual-iso-B'' imaging modality\,\cite{SOM} that allows for faster imaging than the fully quantitative method employed so far. 
Such ``dual-iso-B'' imaging yields a magnetic field-dependent signal\,\cite{SOM}
and with it, an efficient method to determine changes in the PW position. 
We thereby obtain the series of images (3 hours per image) shown in Fig.\,\ref{Fig: Phase Transition}\,\textbf{f} that evidences the stability and incremental motion of the PW through the sample as $B_{\rm ext}$ is decreased. 
This smoothness is further evidence of only weak magnetic pinning in CrSBr. From these data, we extract hysteresis curves (Fig.\,\ref{Fig: Phase Transition}\,\textbf{g}) by determining the relative bilayer areas in the FM and AFM state, as a function of $B_{\rm ext}$\,\cite{SOM}.
The resulting curve is consistent with previous, macroscopic measurements,\,\cite{Ye2022} which supports the notion that past hysteresis measurements in the CrSBr spin-flip transition indeed are explained by PW movement throughout the sample.


\begin{figure}[t]
	\centering
	\includegraphics{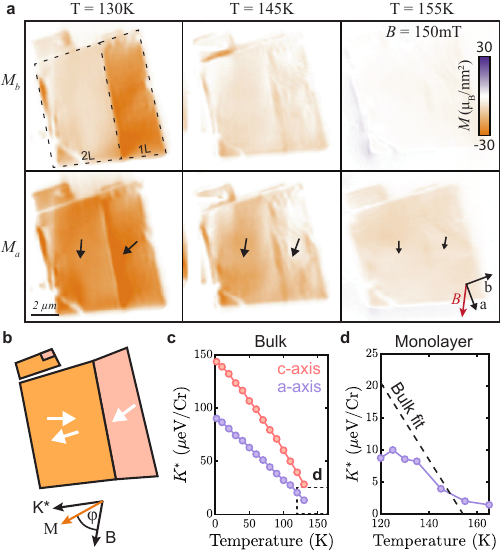}
 \caption{\textbf{Determination of magnetic anisotropy in a CrSBr monolayer.} 
 \textbf{a} Reconstruction\,\cite{SOM, Dubois2022} of magnetization components along the CrSBr $b$-axis ($M_b$, top panels) and $a$-axis ($M_a$, bottom panels) for different temperatures, under a magnetic bias field applied roughly along the $a$-axis ($(\theta, \phi) =  (65^\circ, 265^\circ)$).
\textbf{b} Illustration of the two competing interactions: effective anisotropy ($K^*$) and Zeeman energy ($B$), and their effect on the magnetization direction of the material.
 \textbf{c} Effective in-plane and out-of-plane anisotropy of bulk CrSBr measured with vibrating sample magnetometry\cite{SOM}.
 \textbf{d} Extracted effective in-plane anisotropy ($K^*$) from the measurements in \textbf{a}\cite{SOM} and the extrapolated anisotropy from the bulk measurements in \textbf{c} assuming a linear trend. The dashed box in \textbf{c} indicates the range of \textbf{d}. 
 } \label{Fig:Temp}
\end{figure}

\begin{figure}[ht]
	\centering
	\includegraphics{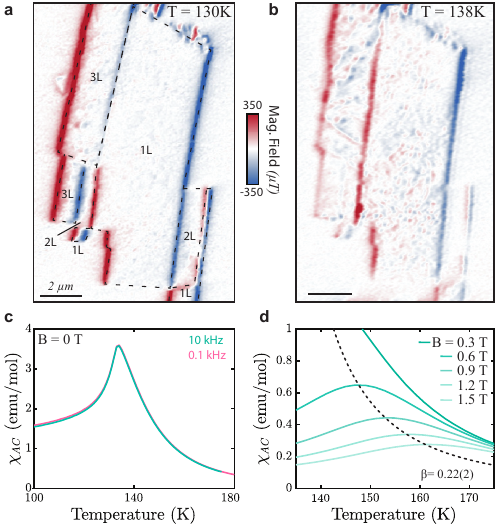}
 \caption{
 \textbf{Properties of CrSBr monolayer near $T=T_N$.} 
 \textbf{a} Magnetic field image of Sample \Stwo at $T = 130~K$ with a bias magnetic field ($B_{\rm ext} = 5~$mT applied along the b-axis). 
 \textbf{b} Same as \textbf{a}, taken at $T = 138~K$. The onset of short-range magnetic inhomogenities is apparent in regions with odd numbers of layers. 
 \textbf{c} Real part of the zero-field, AC magnetic susceptibility $\chi_{AC}$ of bulk CrSBr. 
 The data indicate a lack of additional magnetic phase transitions for $T>T_N=132~$K.
 \textbf{d} Same as \textbf{c} obtained in nonzero bias magnetic fields $B_{\rm ext}$. From these data, we extract critical parameters $\beta = 0.22(2)$ and $\gamma = 2.28(2)$, both approaching the expected values for an ideal 2D-XY spin model ($\beta = 0.231$ and $\gamma = 2.4$\cite{Bedoya2021}). 
 AC susceptibility data were collected with $B$ parallel to the $a$-axis with an oscillating field of $1~$mT. 
 } \label{Fig:2DXYModel}
\end{figure}

We now examine the properties of few-layer CrSBr above and below its magnetic ordering temperature and begin by exploring the material's magnetic anisotropy as a function of temperature. 
For this, we study adjacent mono- and bilayer flakes, where we apply a bias magnetic field $B_{\rm ext}=150~$mT along the intermediate $a$-axis, to tilt the magnetization away from its easy axis, $b$.  
Using our reconstruction method\,\cite{SOM,Dubois2022} we determine the magnetization both along the $a$-axis ($M_a$) and $b$-axis ($M_b$) simultaneously (Fig.\,\ref{Fig:Temp}\,\textbf{a}).
For the bilayer, we observe that the magnetization is aligned with the applied field for all temperatures and steadily decreases with $T$.
This observation is readily explained by spin-canting out of the AFM phase and a decrease of the magnetization as $T$ approaches $T_N$. 

In contrast, the monolayer magnetization for our initial temperature $T=130~$K is initially nearly aligned with the easy axis (Fig.\,\ref{Fig:Temp}\,\textbf{b}), indicating that the effective magnetic anisotropy energy, $K^{*}$, exceeds the Zeeman energy.
As we increase $T$, the monolayer magnetization not only decreases in magnitude but also reorients towards the $a$-axis\cite{SOM}.
This reorientation results from a reduction of $K^{*}$ with increasing temperature due to thermal-magnon induced softening of the ferromagnet\,\cite{Cham2022} (Fig.\,\ref{Fig:Temp}\,\textbf{c}).  
Using the experimentally determined, temperature-dependent magnetization and canting angle of the monolayer flake, we directly and quantitatively extract \textit{K}* for monolayer CrSBr\,\cite{SOM} (Fig.\,\ref{Fig:Temp}\,\textbf{d}). 
Extrapolating the bulk anisotropy measurement shown in Fig.\,\ref{Fig:Temp}\,\textbf{c} to higher temperatures (Fig.\,\ref{Fig:Temp}\,\textbf{d}, dashed line) yields qualitative agreement with our results and indicates that the primary source of magnetic anisotropy for the monolayer is magnetocrystalline anisotropy.

Our data in Fig.\,\ref{Fig:Temp} indicate that FM order in the monolayer persists well beyond the nominal bulk N\'eel temperature $T_N\approx132~$K of CrSBr, consistent with earlier observation of intralayer correlations in CrSBr for $T>T_N$\,\cite{Telford2022,Rizzo2022}. 
To further elucidate the still unclear nature of this intermediate FM regime\,\cite{Lee2021}, we investigate a monolayer CrSBr flake (Fig.\,\ref{Fig:2DXYModel}\,\textbf{a}) in a temperature range slightly above $T_N$\footnote{We note that due to microwave heating the temperature measured constitutes a lower bound for the sample temperature, which is likely a few Kelvin higher than the indicated temperature values -- see SI for more details.} and at low magnetic fields.
We observe that once $T\gtrsim T_N$, the monolayer develops sizable spatial variations of magnetization (Fig.\,\ref{Fig:2DXYModel}\,\textbf{b}) that are not observed once $T\lesssim T_N$
\footnote{We note the spatial gradient in the magnetization variation that is visible in Fig.\,\ref{Fig:2DXYModel}\,\textbf{b}. 
We tentatively assign this gradient to a gradient in temperature resulting from heating due to the microwave control line, used to drive the NV, that was located on-chip and near the left edge of the image shown in Fig.\,\ref{Fig:2DXYModel}\,\textbf{b}.}.
We also observed such inhomogeneities on trilayer flakes, whereas even-numbered layers do not exhibit any noticeable inhomogeneity (see bilayer sections in Fig.\,\ref{Fig:2DXYModel}\,\textbf{a,\,b} and additional data\,\cite{SOM}).
The latter suggests that weak remaining AFM interlayer coupling above $T_N$ can still compensate for the observed inhomogeneities. 

Further experimental observations help us identify the most likely origin of the inhomogeneities in the intermediate FM phase.
First, we note that the inhomogeneities appear completely static over the timescale of our experiments of several days.
This excludes critical spin fluctuations near $T_c$, or the appearance of unbound meron/anti-meron pairs\,\cite{Moro2022} as their origin.
Second, a near-homogenous monolayer magnetization can be restored by increasing $B_{\rm ext}$ to around $200~$mT\cite{SOM}, indicating that the saturation magnetization is nearly constant across the sample.
Given that the magnetic anisotropy $K^*$ is near-vanishing for $T\gtrsim T_N$ variations in $K^*$ appear unlikely as the origin for the observed inhomogeneities. 
This would leave local variations of intralayer exchange couplings as the most likely explanation for our observation.
These local variations could originate from atomistic defects, strain variation, or other local disorders that could disrupt the regular crystallographic structure.

To further elucidate the nature of magnetic ordering in the intermediate magnetic phase of CrSBr, we turn to measurements on bulk crystals.
In Fig.\,\ref{Fig:2DXYModel}\,\textbf{c}, we show the real part
\footnote{The imaginary part of $\chi_{AC}$ remains $\approx 0$ over the entire temperature range studied here} 
of the zero-field AC susceptibility $\chi_{AC}$ of a bulk CrSBr crystal.
Within the probed parameter range, $\chi_{AC}$ is frequency independent and displays a single cusp at $T_N=132~$K. 
The absence of an additional higher temperature feature in $\chi_{AC}$ indicates that, despite the presence of FM correlations persisting to temperatures $T>T_N$, no short- or long-range intralayer order emerges in CrSBr above $T_N$ at zero-field.
However, in the presence of applied bias magnetic fields, a field-dependent maximum in $\chi_{AC}(T)$ emerges   (Fig.\,\ref{Fig:2DXYModel}\,\textbf{d}). 
The magnetic field dependence of this high-temperature feature can be used to extract critical exponents for magnetic ordering in CrSBr\cite{Zhao1999, Stanley1987}.
Intriguingly, the extracted values of $\beta = 0.22(2)$ and $\gamma = 2.28(2)$ are close to the values expected for the 2D-XY model ($\beta = 0.231$ and $\gamma = 2.4$\cite{Bedoya2021,Bramwell1993}).
The value of $\beta$ we determined aligns with values previously determined by a range of experimental techniques\,\cite{Scheie2022,Lopez2022} and earlier conclusions that CrSBr follows 2D-XY-like behavior above $T_N$\,\cite{Moro2022}.

Our combined findings indicate that easy-plane behavior of decoupled FM layers (e.g. 1L, 3L, ..., etc) could exist in a narrow temperature range above $T_N$ and below $T_C$, in which 2D-XY-like physics could thereby be observed in CrSBr. 
However, in our present samples, this behavior is likely masked by the inhomogeneities in intralayer exchange couplings, as discussed earlier. 
Future advances in material purification\,\cite{telford_designing_2022} and isolation from the substrate could therefore offer exciting perspectives for observing 2D-XY spin physics\,\cite{kosterlitz1973} with its accompanying topological spin textures\,\cite{Moro2022} in monolayers of the CrSBr family.


In this work, we provided a quantitative, nanoscale study of magnetization strength and anisotropy in few layer CrSBr samples and addressed two key magnetic phase transitions by direct magnetic imaging using single-spin magnetometry. 
We observed that the magnetization per CrSBr layer decreases monotonically with layer number, but remains nonzero even for monolayers, which we find magnetically stable even in the absence of encapsulation.
We further investigated the AFM-FM spin-flip transition in bilayers and found them to be driven by the nucleation and subsequent propagation of a phase wall, rather than a coherent rotation of layer magnetization.
Finally, we addressed the evolution of magnetization near CrSBr's critical temperature and directly evidenced the reduction of anisotropy when approaching $T_N$. 
Near $T_N$, we observed the onset of magnetic inhomogeneities in odd-layer flakes, which we attributed to local disorder in intralayer exchange couplings that currently mask a native 2D-XY behaviour of CrSBr monolayers.

Our results underline CrSBr's significant potential for the development of novel technologies based on 2D magnets.
In particular, the air stability, and large-range uniformity of magnetization across tens of microns evidence robustness and scalability of this material, while the highly stable magnetic PWs we discovered suggest potential interesting functionalities in the context of spintronics and racetrack memory devices\cite{Parkin2008}. 

\small

\textbf{Author contributions}
The NV measurements were performed by MAT and DAB together with CS and PR, under the supervision of PM. DAB, MAT and CS analysed the Data. The Samples were prepared by EJT and JC, and the bulk crystal was synthesised by DGC, all under the supervision of XR and CRD. 
MEZ performed bulk AC magnetic measurements and data analysis.
BG and MP provided micromagnetic simulation of the material while RRE and EJGS provide theoretical description of the formation of the phase-boundary. 
DAB, MAT and PM wrote the manuscript.
All authors discussed the data and commented on the manuscript.

\textbf{Acknowledgment}
We acknowledge financial support by the ERC consolidator grant project QS2DM, 
by SNF project No. 188521, and from the National Centre of Competence in Research (NCCR) Quantum Science and Technology (QSIT), a competence centre funded by the Swiss National Science Foundation (SNF).
Synthesis of the CrSBr crystals was funded by the Columbia MRSEC on Precision-Assembled Quantum Materials (PAQM) under award number DMR-2011738 and the Air Force Office of Scientific Research under grant FA9550-22-1-0389. Bulk magnetic measurements were supported under Energy Frontier Research Center on Programmable Quantum Materials funded by the US Department of Energy (DOE), Office of Science, Basic Energy Sciences (BES), under award DE-SC0019443. The instrument used to perform these magnetic measurements was purchased with financial support from the National Science Foundation through a supplement to award DMR-1751949.
EJGS acknowledges computational resources through CIRRUS Tier-2 HPC Service (ec131 Cirrus Project) at EPCC (http://www.cirrus.ac.uk) funded by the University of Edinburgh and EPSRC (EP/P020267/1); ARCHER UK National Supercomputing Service (http://www.archer.ac.uk) {\it via} Project d429. E.J.G.S. acknowledges the EPSRC Open Fellowship (EP/T021578/1), and the Edinburgh-Rice Strategic Collaboration Awards for funding support.
Micromagnetic calculations were performed at sciCORE (http://scicore.unibas.ch/) scientific computing center at University of Basel.
B.G. acknowledges the support of the Canton Aargau.

\bibliographystyle{apsrev4-2}

\bibliography{Main_CrSBr_Arxiv}


\clearpage
\pagebreak
\newcommand{\beginsupplement}{%
	\setcounter{table}{0}
	\renewcommand{\thetable}{S\arabic{table}}%
	\setcounter{figure}{0}
	\renewcommand*{\thefigure}{S\arabic{figure}}%
}

\beginsupplement
\onecolumngrid

\begin{center}
	\large
	\textbf{Supplementary information for\\ "Nanoscale magnetism and magnetic phase transitions in atomically thin CrSBr''}
\end{center}

\normalsize

\maketitle

\tableofcontents
\newpage

\section{Sample fabrication}\label{Sec: Sample Fab}

\textbf{Synthesis of CrSBr bulk crystals:}
Large single crystals of CrSBr were grown using a chemical vapor transport reaction described in ref.~\cite{Scheie2022}.

\textbf{Fabrication of non-encapsulated CrSBr samples:}
CrSBr flakes were exfoliated onto 90~nm SiO$_2$/Si$^+$ substrates (NOVA HS39626-OX9) using mechanical exfoliation with Scotch\textsuperscript{\textregistered} Magic$^{\rm TM}$ tape~\cite{Novoseloc2004, Novoselov2005}. 
Before exfoliation, the substrates were cleaned with a gentle oxygen plasma to remove adsorbates from the surface and increase flake adhesion~\cite{Huang2015}. 
The exfoliation was done under ambient conditions. Flake thickness was identified using optical contrast and then confirmed with atomic force microscopy~\cite{Telford2022, Lee2021}.

\textbf{Fabrication of encapsulated CrSBr samples:}
CrSBr flakes were exfoliated onto 285~nm SiO$_2$/Si$^+$ substrates (NOVA HS39626-OX) using mechanical exfoliation with Scotch\textsuperscript{\textregistered} Magic$^{\rm TM}$ tape~\cite{Novoseloc2004, Novoselov2005}. 
Before exfoliation, the substrates were cleaned with a gentle oxygen plasma to remove adsorbates from the surface and increase flake adhesion~\cite{Huang2015}. 
The exfoliation was done under inert conditions in an N$_2$ glovebox with <~1~ppm O$_2$ and <~1~ppm H$_2$O content. 
Thin flakes (<~10~nm thick) of hexagonal boron nitride (hBN) were then placed on top of the desired CrSBr flakes to encapsulate them using the dry-polymer transfer technique~\cite{Wang2013}. 
CrSBr flake thickness was identified using optical contrast and then confirmed with atomic force microscopy after encapsulation with hBN~\cite{Telford2022, Lee2021}.

\textbf{Atomic force microscopy for sample screening:}
Atomic force microscopy was performed in a Bruker Dimension Icon\textsuperscript{\textregistered} using OTESPA-R3 tips in tapping mode. 
Flake thicknesses were extracted using Gwyddion to measure histograms of the height difference between the substrate and the desired CrSBr flake.

\section{NV magnetometry Experimental information}

\subsection{Apparatus description}\label{Sec: NV apparatus}

\begin{figure*}
	\centering
	\includegraphics{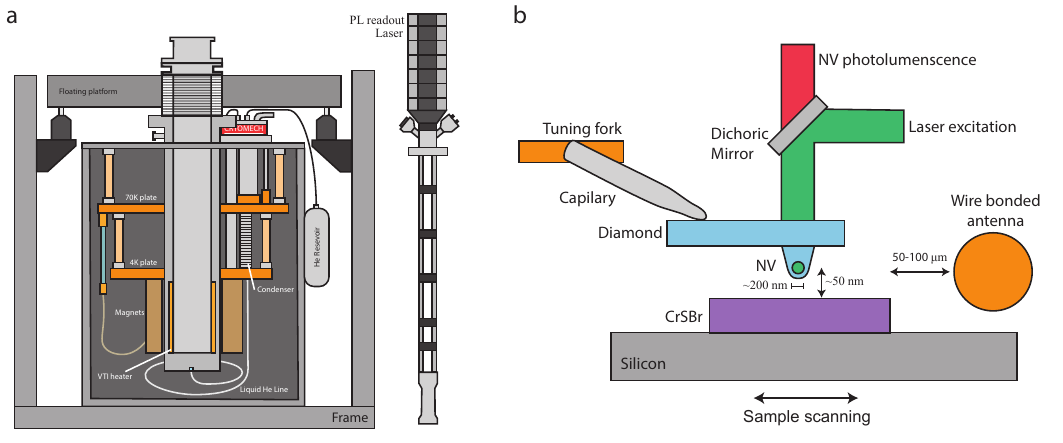}
	\caption{Experimental principle and setup. 
		\textbf{a} Illustration of the Atto2200 Dry system that consists of the variable temperature insert (VTI) and a standard attocube AFM-CFM microscope, with a diamond AFM cantilever (Qnami) for magnetometry. 
		\textbf{b} Illustration of the diamond AFM that is brought into contact with the sample that is scanned to image the material. Microwave control is applied via a wire that is bonded across the sample. }
	\label{SIFig: exp}
\end{figure*}

The magnetic images were taken using two different Attocube cryostats, an attoLIQUID 1000 for measurements around $T=4$K and an attoDRY 2200 for variable temperatures $T = 2 - 300$K measurements (see Fig.~\ref{SIFig: exp}a). 
Each system has a similar control apparatus, see Figure \ref{SIFig: exp}b for details.
The NV electronic spin is excited using a 532 nm laser (attoLiquid - Laser Quantum 532 Gem, and attoDry - Laser Quantum 532 Torus) and then the photoluminescence (PL) of the NV spin is separated using a dichroic mirror which is detected with an avalanche photodiode (Excelitas SPCM-AQRH-33). 
The microwave control is performed using a signal generator (attoLIQUID - SRS SG384, attoDry - Rhode Schwartz SMBV100B) which is applied to the NV using a wire that is bonded across the sample.
To maintain the optical readout of the NV spin, the sample is scanned while the diamond tip remains static. 

The NV spin has a standoff from the sample of approximately 50 nm which is a combination of the depth of the NV in the diamond tip and the additional standoff given by the broad AFM tip ($\sim 200$ nm diameter at the end of the tip). 
This results in a standoff limited resolution of the magnetic images that varies from 30 to 100~nm depending on the tip, but is typically in the range of 50~nm.

\subsection{Measurement techniques}\label{Sec: Measurement techniques}

\begin{figure*}
	\centering
	\includegraphics{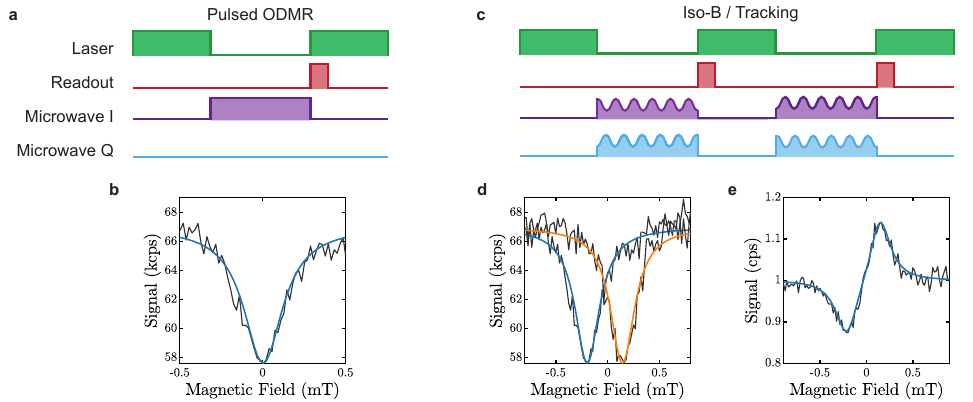}
	\caption{\textbf{Measurement techniques.} 
		\textbf{a} ODMR pulse sequence where the NV is initialised using a 532 nm laser pulse, followed by MW control pulse to transfer the spin into a darker state, and finally the spin is readout using another laser pulse. The continous wave version of this sequence has the laser, MW, and readout on all of the time. 
		\textbf{b} Example ODMR spectrum (black) and Lorentzian fit (blue) taken using the sequence in \textbf{a} by sweeping the MW frequency across the resonant transition.
		\textbf{c} Pulse sequence for frequency modulated tracking, where an IQ-modulator is used to combine frequencies such that the applied frequency is shifted between each readout by a frequency $\Delta f$. 
		\textbf{d} Example ODMR spectra using the frequency modulation (black) with Lorentzian fits (blue and orange). 
		\textbf{e} Normalised signal from \textbf{d} that is used for both single point Iso-B imaging and frequency tracking where a PID loop maintains a normalised signal of 0.     }
	\label{SIFig: measure}
\end{figure*}

Various NV magnetometry techniques are used in this work. 
The primary is a pulsed or continuous wave optically detected magnetic resonance (ODMR). This sequence uses a laser pulse to initalise the NV into the \ket{0} state. 
Then during the free evolution time a $\pi$-rotation microwave (MW) pulse is applied when on resonance with the transition. 
Finally, a second laser pulse is used to optically read out the NV state and reinitialise the NV back into the \ket{0} state. 
Full sequence is shown in Fig.~\ref{SIFig: measure}~\textbf{a} and example spectrum shown in Fig.~\ref{SIFig: measure}~\textbf{b}. 
To extract the central NV frequency when using this method, each pixel is fitted with a Lorentzian function.

An alternative measurement technique involves frequency modulation of the MW to tune the applied frequency rapidly. 
This involves modulating the applied frequency using an IQ modulator where the frequency is shifted to a lower frequency using the functions:
\begin{align}
I &= \cos (t \Delta f /2) \\
Q &= \sin(t \Delta f /2)	
\end{align}   
and to higher frequency
\begin{align}
	I &= \sin (t\Delta f/2) \\
	Q &= \cos(t\Delta f/2)	
\end{align} 
where $\Delta f$ is the desired splitting of the two peaks which is typically set to the width of the ODMR feature and $t$ is the time. 
The full sequence is shown in Fig.~\ref{SIFig: measure}~\textbf{c}. 
This sequence results in two ODMR features that are centred around the true frequency as shown in Fig.~\ref{SIFig: measure}\textbf{d}. 
By normalising these two signals such that
\begin{eqnarray}
	S = \frac{\text{ODMR}_1}{\text{ODMR}_2}
\end{eqnarray}
we obtain a single spectrum that can be locked onto as shown in Fig.~\ref{SIFig: measure}\textbf{e}.

In this work, we use this signal in two different ways. 
The first is to perform a single frequency measurement, referred to as Iso-B. 
This technique sets the applied frequency such that when away from the sample magnetic field the signal $S = 0$. 
Then local changes in magnetic field are measured as changes in this signal. 
This allows for fast measurements but has the cost that the dynamic range is limited to the linear regime of the signal, which in this case is approximately $500~\mu$T. 
Beyond this, the magnetic fields can be measured but are not quantitative. 

The other technique is to lock onto the zero signal point using a PID loop. 
In this fashion, the central frequency of the NV is tracked throughout the measurement. 
To minimise, effects from the PID loop lagging behind the true signal this technique requires a longer integration time per point than the Iso-B technique but provided the magnetic field gradient is less than the width of the signal, this technique has an unlimited dynamic range.

\section{Sample characterisation}
In this work, we studied various different CrSBr samples, as summarised in table~\ref{Tab: Samples}. 
The samples contain different layer thicknesses of CrSBr ranging from monolayer to five layers. 
Samples \Sone, \Sthree and \Sfour were not encapsulated and exposed to air for several hours to days before measurements. Sample \Stwo was encapsulated with hBN. 
All the samples were characterised optically and by scanning NV magnetometry measurements as described in the next sections.

\begin{table}[h]
	\centering
	\bgroup
	\def\arraystretch{1.5}
	\begin{tabular}{ c | c | c | c}
		Sample \#~ & ~Label~ & ~Layers~ & ~Encapsulated~ \\
		\hline
		 1 & \Sone   & 2, 3, 5 		& No \\ 	
		 2 & \Stwo & 1, 2, 3 		& Yes \\ 	
		 3 & \Sthree & 1, 2, 4, 5 		& No \\ 	
		 4 & \Sfour  & 1, 2 			& No  	
	\end{tabular}
	\egroup
	\caption{Description of samples used in this study.}
	\label{Tab: Samples}
\end{table}

\subsection{Optical sample characterisation}

\begin{figure}[h]
	\centering
	\includegraphics{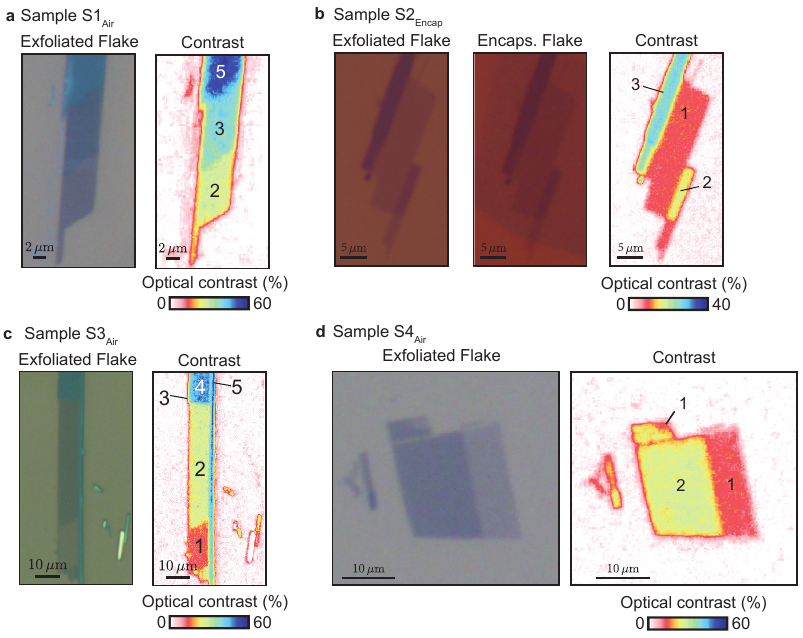}
	\caption{Optical sample characterisation of the CrSBr crystals studied in the main text and labelled accordingly. 
		Panels \textbf{a} to \textbf{d} show the optical microscope images of samples \Sone to \Sfour and the extracted optical contrast used to determine the layer thicknesses. 
		For the sample~\Stwo the encapsulated flake as depicted as well in \textbf{b}.
	}
	\label{SIFig: optical sample charc}
\end{figure}

To assess the layer composition of the CrSBr flakes studied in this work (see table~\ref{Tab: Samples} for sample overview) we optically characterised the individual samples as shown in Fig.~\ref{SIFig: optical sample charc}. 
The layer thickness of the CrSBr samples was determined by the relative optical contrast between the CrSBr flakes and a SiO2 layer. 
A single layer of this material has an absorption of approximately 15\% in our optical microscope. The absorption scales linearly for thin layers (<~3) and is less accurate for thicker layers (>~5). 
For the encapsulated flakes of sample \Stwo additional AFM measurements were taken of the flake after encapsulation to independently verify the layer thicknesses.

\subsection{Layer dependent magnetisation}

\begin{figure}
	\centering
	\includegraphics{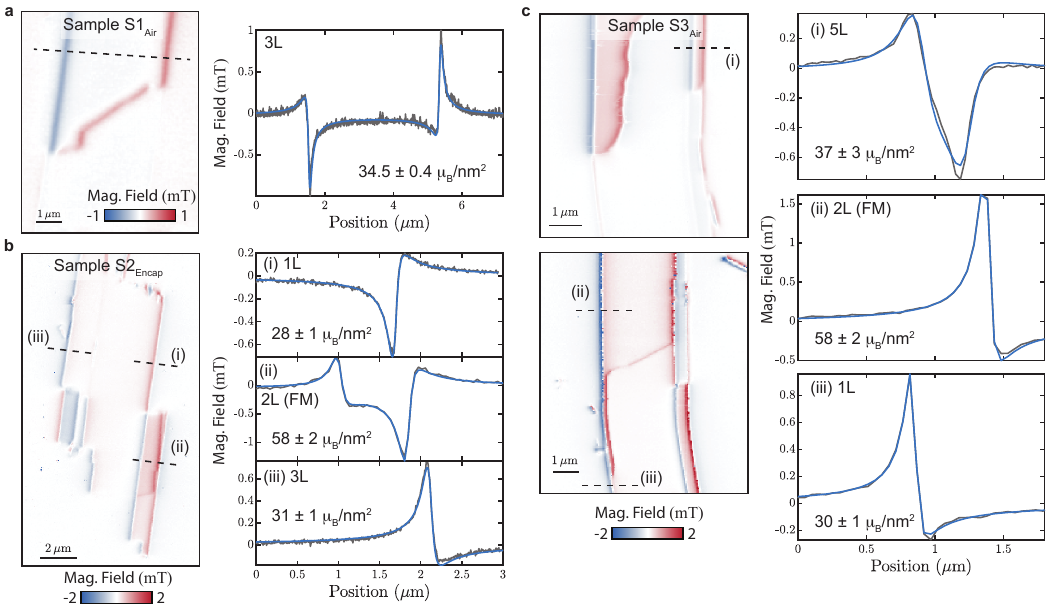}
	\caption{Fitted Linecuts of the magnetic field for layer dependent magnetisation determination. 
		\textbf{a} Magnetic image and linecut across the tri-layer for sample \Sone.
        \textbf{b} Magnetic image and linecuts from sample \Stwo for mono, bi, and tri-layer sections.
        \textbf{c}  Magnetic image and linecuts from sample \Sthree for mono, bi, and 5-layer sections.
	}
	\label{SIFig: Linecuts}
\end{figure}

\begin{figure}
    \centering
    \includegraphics{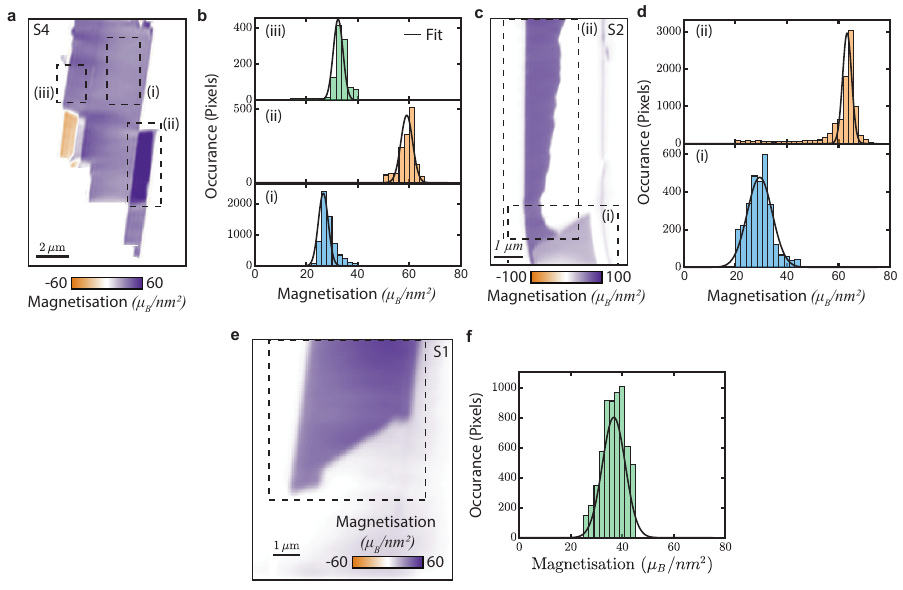}
    \caption{Histograms of the magnetisation reconstruction for layer dependent magnetisation determination. 
    \textbf{a} Magnetisation reconstruction for sample \Sfour with regions for where histograms are taken. 
    \textbf{b} Histograms from \textbf{a}, where blue is monolayer, orange is bilayer, and green is trilayer. The fit to each histogram is shown in black. 
    \textbf{c} Magnetisation reconstruction for sample \Stwo
    \textbf{d} as in \textbf{b} but for the magnetisation of sample \Stwo in \textbf{c}.
    \textbf{e} Magnetisation reconstruction for sample \Sone
    \textbf{f} as in \textbf{b} but for the magnetisation of sample S1 in \textbf{e}.
    }
    \label{SIFig: Histograms}
\end{figure}

To determine the magnetisation of the CrSBr flakes we use two separate methods: The first method is fitting the magnetic stray field of a 1D line cut across the edge of a flake and the second method is taking a histogram of the reconstructed magnetisation image of a CrSBr flake.
In the following paragraph we describe the two methods for extracting the magnetisation of the studied CrSBr and we summarised all the results in table~\ref{Tab: Magnetisation values}.

The first method we used to extract magnetisation from our measurements is by fitting the magnetic field of a 1D line across the edge of the material. 
The magnetic field from a stray edge with an arbitrary magnetisation direction is given by:
\begin{align}
    B_x &= - \frac{\mu_0 M}{2 \pi}  \frac{z_{NV} \cos(\theta_M) + (x - x_0) \sin(\theta_M)\cos(\phi_M) }{z_{NV}^2 + (x - x_0)^2 },  \\
    B_z &= \frac{\mu_0 M}{2 \pi} \frac{ - z_{NV} \sin(\theta_M) + (x - x_0) \cos(\theta_M)\cos(\phi_M)}{z_{NV}^2 + (x - x_0)^2 },
\end{align}
where $\mu_0$ is the vacuum permeability, $M$ is the magnetisation magnitude of the material, $z_{NV}$ is the stand-off distance from the NV to the sample, $x$ is the position vector and $x_0$ is the position of the sample edge, $\theta_M$ is the angle of the magnetisation from the z-axis,  and $\phi_M$ is the azimuthal angle of the magnetisation. 
The values for the NV angles $\theta_M$ and $\phi_M$ were independently determined in our experiments and are kept fixed during the fitting routine. 
Examples of these linecuts and their respective fits are shown in Fig.~\ref{SIFig: Linecuts}. The extracted magnetisation and NV stand off distance are summarised in table~\ref{Tab: Magnetisation values}.

The second method of determining the magnetisation is to use a neural network to reconstruct a magnetisation image (method explained later in Section~\ref{Sec: reconstruction}) and then take a histogram of the image to determine the average magnetisation of a given region. 
Examples of these histograms are shown in Fig.~\ref{SIFig: Histograms}.

For many of the samples, we were able to obtain estimations of the magnetisation with one or both of the methods for the different number of layers. 
The results of which are shown in Table~\ref{Tab: Magnetisation values}. 
Comparing these results shows that both methods for determining the magnetisation give a similar value of magnetisation for a specific layer thickness of CrSBr.


\begin{table}[h]
	\centering
	\bgroup
	\def\arraystretch{1.5}
	\begin{tabular}{ c | c | c | c | c | c | c | c }
		Layers~ & ~Sample~ & State & ~$M$ ($\mu_b/$nm$^2$)~ & ~$M$ per layer~  & $B$ (mT) & Height (nm) & ~Method  \\
		\hline 
		1 & \Stwo & AFM & 28 $\pm$ 1 & 28 $\pm $ 1 & 150 & 63 $\pm$ 1 & Line cut  \\ 
		1 & \Stwo & AFM & 29 $\pm$ 2 & 29 $\pm$ 2 & 150 & 66 $\pm$ 4 & Line cut \\ 
        1 & \Stwo & AFM & 28 $\pm$ 3 & 28 $\pm$ 3 & 150 & 70 & \color{blue}Histogram \\
		1 & \Sthree & AFM & 30 $\pm$ 1 & 30 $\pm$ 1 & 230 & 48 $\pm$ 2 & Line cut \\ 
		1 & \Sthree & AFM & 27 $\pm$ 2 & 27 $\pm$ 2 & 150 & 60 & Histogram \\
  	1 & \Sthree & AFM & 30 $\pm$ 6 & 30 $\pm$ 6 & 230 &       60 & \color{blue}Histogram \\ 
        \hline 
        2 & \Stwo & FM & 58 $\pm$ 2 & 29 $\pm$ 1 & 150 & 70 & Line cut \\ 
        2 & \Stwo & FM & 58 $\pm$ 3 & 29 $\pm$ 1.5 & 150 & 70 & \color{blue}Histogram \\ 
		2 & \Sthree & FM & 64 $\pm$ 3 & 32 $\pm$ 1.5 & 180 & 55 $\pm$ 2 & Line cut \\ 
        2 & \Sthree & FM & 61 $\pm$ 8 & 30.5 $\pm$ 4 & 170 & 100 & Histogram \\
		2 & \Sthree & FM & 63.3 $\pm$ 0.2 & 31.6 $\pm$ 0.1 & 180 & 60 & \color{blue}Histogram \\
		2 & \Sthree & FM & 58 $\pm$ 7 & 29 $\pm$ 4 & 230 & 60 & Histogram \\ 
		2 & \Sthree & FM & 58 $\pm$ 2 & 29 $\pm$ 1 & 230 & 46 $\pm$ 2 & Line cut \\ 
        \hline 
		3 & \Sone & AFM & 34.5 $\pm$ 0.4 & 34.5 $\pm$ 0.4 & 150 & 63 $\pm$ 1 & Line cut \\ 
		3 & \Sone & AFM & 36.7 $\pm$ 0.2 & 36.7 $\pm$ 0.2 & 5 & 60 & \color{blue}Histogram \\ %
		3 & \Stwo & AFM & 31 $\pm$ 0.7 & 31 $\pm$ 0.7 & 150 & 69 $\pm$ 2 & Line cut \\ 
		3 & \Stwo & AFM & 33 $\pm$ 3 & 33 $\pm$ 3 & 150 & 70 & \color{blue}Histogram \\ %
		3 & \Stwo & FM & 98 $\pm$ 5 & 33 $\pm$ 2 & 300 & 73 $\pm$ 5 & Line cut \\ 
        \hline 
		5 & \Sthree & AFM & 37 $\pm$ 3 & 37 $\pm$ 3 & 172.5 & 111 $\pm$ 10 & Line cut \\ 
  		5 & \Sthree & AFM & 38 $\pm$ 3 & 38 $\pm$ 3 & 170 & 100 & Histogram \\ 
    5 & \Sthree & AFM & 37 $\pm$ 2 & 37 $\pm$ 2 & 160 & 100 & Histogram \\ 
	\end{tabular}
	\egroup
	\caption{Measurements of the magnetisation of different layer thicknesses at a temperature of approximately 4 K. Blue color highlights data points used in Fig.1\,e of the main text. Note: we did not acquire any data at low temperature of \Sfour that was suitable for extracting the magnetisation. }
	\label{Tab: Magnetisation values}
\end{table}

\section{AFM to FM phase transition}

In this section, we give additional details on the measurements of the AFM to FM phase transition in the main text. 
We also show additional data that was not used in the main text. 

\subsection{Phase boundary movement}

\begin{figure*}
	\centering
	\includegraphics{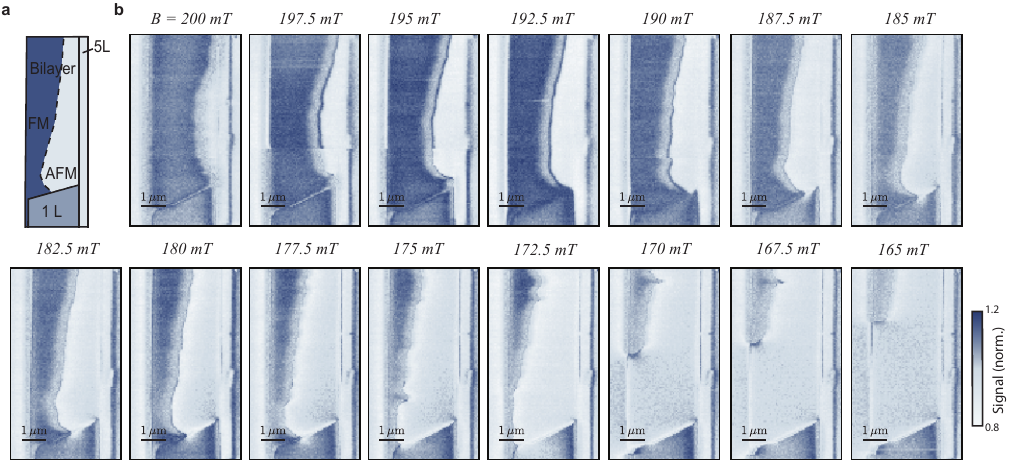}
	\caption{Measurement series of phase boundary movement through bilayer CrSBr going from high field (200~mT) to low field (165~mT). 
		\textbf{a} Illustration of bilayer CrSBr with FM (dark blue) and AFM (light blue) region.
		\textbf{b} Measurements were taken with single point Iso-B imaging, method described in Section~\ref{Sec: Measurement techniques}.}
	\label{SIFig: domain movement}
\end{figure*}

As mentioned in the main text, we imaged the phase boundary movement through the bilayer using the Iso-B technique described in Section \ref{Sec: Measurement techniques}. 
The applied frequency was tuned to give a signal of zero while in contact away from the flake, resulting in more magnetised regions having a higher signal value. 
To perform these sweeps we first applied a large enough field ($B = 300$ mT) to fully initialise the bilayer into the FM state. 
We then decreased the field incrementally, taking an image (approximately 3 hours) at each field strength. 

In Fig.~\ref{SIFig: domain movement} we show an additional dataset of the same flake but from a separate measurement series to that shown in the main text. 
The image region consistences of a bilayer and monolayer section with an adjoining thicker sections (4L) that remains in the AFM state throughout the series (Fig.~\ref{SIFig: domain movement}~a).
As the field is decreased the FM region in the bilayer recedes  (Fig.~\ref{SIFig: domain movement}~b).
During this process, the FM phase boundary moves faster along the a-axis (intermediate axis) of the material than the b-axis (easy axis). 
The tendency for the phase wall to move along the a-axis was also observed in a trilayer flake which will be discussed in Section~\ref{Sec: trilayer FM}.

\subsection{Formation of FM region in trilayer CrSBr} \label{Sec: trilayer FM}

\begin{figure*}
	\centering
	\includegraphics{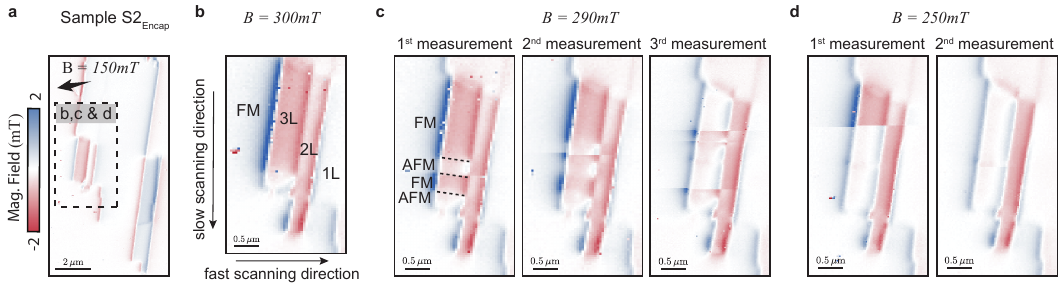}
	\caption{Measurement series of the formation of FM/AFM regions in trilayer CrSBr.
    \textbf{a} Magnetic field map of sample~\Stwo.
    Dashed box marks the region of the trilayer region shown in panels \textbf{b-d}.
    \textbf{b} Magnetic field map of FM ordered trilayer CrSBr measured in an external field of $B=300$\,mT. 
    \textbf{c} Consecutive measurements of the formation of AFM regions in predominantly FM ordered trilayer CrSBr.
    The flake was first initialised in its FM state by applying $B > 300$\,mT and subsequently measured in a lower field (290mT) for three times in a row (1st to 3rd measurement). 
 \textbf{d} Analog to \textbf{c} but measured in a lower external B-field (250mT).
 }
	\label{SIFig: domain trilayer}
\end{figure*}

To investigate the domain movement in trilayer CrSBr, we imaged a trilayer section of sample~\Stwo, see Fig.~\ref{SIFig: domain trilayer}a. The small trilayer section flipped to a FM state in an external magnetic field of $B = 300$~mT applied along the NV axis, Fig.~\ref{SIFig: domain trilayer}b. Reducing the applied field to $B = 290$~mT introduced AFM domains in the trilayer flake, Fig.~\ref{SIFig: domain trilayer}c. These domains were evolving as a function of time: by re-imaging the flake at the same field ($B = 290$~mT) the AFM regions extended further. Reinitilaizing the trilayer flake in the FM state by applying $B = 300$~mT and subsequently imaging the flake at $B = 250$~mT shows a similar trend: AFM domains form and propagate though the flake, Fig.~\ref{SIFig: domain trilayer}d. In both cases, the domains propagated along the a-axis (intermediate axis) of the CrSBr flake. This finding is consistent with the domain movement observed in bilayer CrSBr, Fig.~\ref{SIFig: domain movement}, where propagation along the a-axis was preferred compared to the b-axis (easy axis). Differently to the domain in bilayer CrSBr, which was stable for multiple hours under an external applied magnetic field, the domain observed in trilayer was changing during scans, most likely due to a more meta-stable state of the FM/AFM phase coexistence or an effect of dragging the domain with laser light. 

\subsection{Hysteresis loop}

\begin{figure*}
	\centering
	\includegraphics{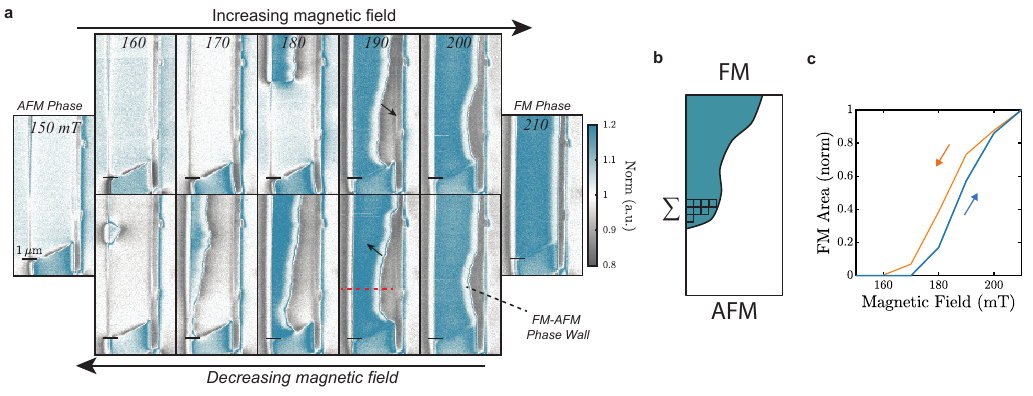}
	\caption{Measurements of domain wall hysteresis. 
	\textbf{a} Measurement series of the domain wall movement through the material using the Iso-B technique for both increasing and decreasing fields. 
\textbf{b} Illustration of partially flipped state with thresholding to sum up the pixels in the ferromagnetic state.
\textbf{c} Magnetic hystersis extracted from the images in \textbf{a}.   }
	\label{SIFig: Hysteresis}
\end{figure*}

Using the Iso-B measurement method we took a series of images at different magnetic fields to map out the hysteresis of the domain wall movement through the bi-layer, see Fig.~\ref{SIFig: Hysteresis}\textbf{a}. These images show a clear difference in the domain wall position with different magnetic sweeping directions. In order to calculate the hysteresis of the domain wall we sum the number of pixels that are in the ferromagnetic state, as illustrated in Fig.~\ref{SIFig: Hysteresis}\textbf{b}. Where the ferromagnetic state is define as pixels with a normalised signal greater than $1.05$. This is this normalised by the number of total pixels on the bi-layer region, such that
\begin{equation}
	M = \frac{\sum \text{pixels}_{FM}}{\sum \text{pixels}}.
\end{equation}
 While is not a perfect estimation of the region it roughly approximates that FM region. The final results is than normalised by the state when the flake to purely in the FM state to force this value to be 1 and minimise issues with the imperfect threshold definition. This results in the hysteresis curve shown in Fig.~\ref{SIFig: Hysteresis}\textbf{c}.

\section{Temperature dependence}\label{Sec: Temperature}

A series of different measurements were performed at elevated temperatures with the aim of extracting the in-plane anisotropy of the material and investigating the loss of long range order. This is discussed in detail below. 

In all measurements (unless otherwise stated) the temperature was measured by a resistivity thermal couple located at the top of the piezo stack which held the sample. Due to the application of microwaves to drive the NV spin we expect some offset in the real sample temperature compared with that sensor temperature. 
This offset is MW power dependent and hard to calibrate for a large range of temperatures. 
As such, we reference the temperature that was measured by the sensor, which ultimately means the real temperature is several unknown degrees warmer. 

\subsection{Anisotropy measurement series}

\begin{figure*}
	\centering
	\includegraphics{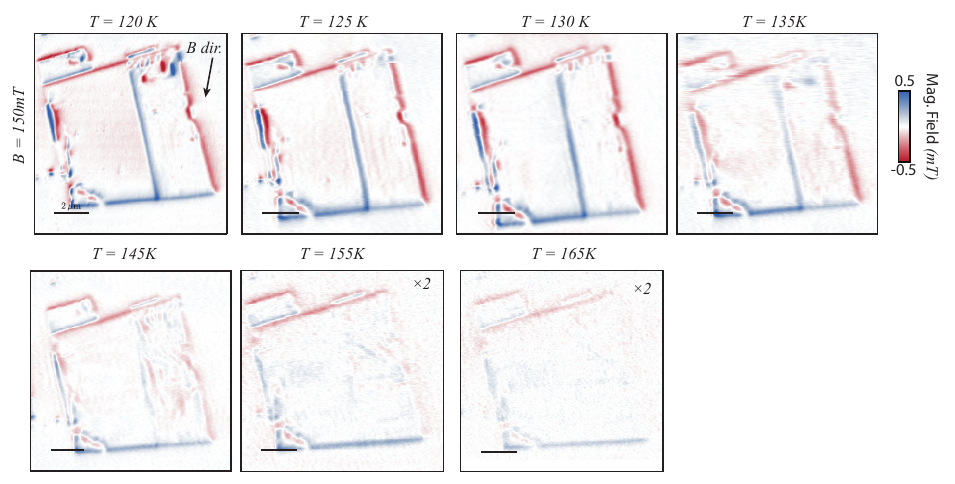}
	\caption{Temperature series of monolayer and bilayer sample. Full temperature series of sample \Sfour shown in the main text all taken with a magnetic field strength of $B = 150 mT$ roughly along the a-axis of the material with an out of plane component of approximately $28^\circ$.}
	\label{SIFig: Temp series}
\end{figure*}

To investigate the transition from AFM to FM at higher temperatures we applied a magnetic field along the intermediate a-axis of \Sfour. In doing so we partially magnetised the bilayer to point along the B-field while the monolayer was still anisotropy dominated, resulting in it pointing along the b-axis of the material. 
We then maintained this magnetic field ($B = 150$ mT, $\phi = 262^\circ$, and $\theta = 62^\circ$) while increasing the temperature, imaging the magnetic field at each temperature. 
The magnetic images are all shown in Fig.~\ref{SIFig: Temp series}. 
This data is used for the reconstruction of the non-uniform magnetisation described in Section~\ref{Sec: reconstruction} and the extraction of in-plane anisotropy in the main text.  

\subsection{Emergence of the near 2D-XY interaction model}

We took a series of measurements on different samples to investigate the emergence of the high temperature near 2D-XY regime, referred to here as the anisotropy transition. 

\begin{figure*}
	\centering
	\includegraphics{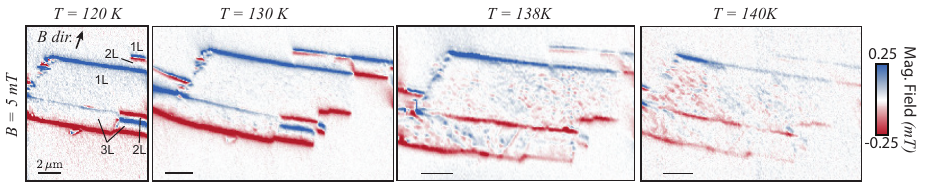}
	\caption{Temperature series of sample D5. Full temperature series of the sample shown in the main text all taken with a magnetic field strength of $B = 5 mT$ roughly along the b-axis of the material with an out-of-plane component of approximately $35^\circ$.}
	\label{SIFig: Temp series 2}
\end{figure*}

\textbf{Anisotropy transition in encapsulated material.}
In the main, text we showed magnetic images of sample \Stwo at elevated temperatures. 
This measurement series was taken similar to the anisotropy series. 
We initialised the flake through cooling below the critical temperature with zero magnetic field. 
At $T = 120$~K we applied a small magnetic field $B = (\theta, \phi, \left|B\right|) = (55^\circ, 80^\circ, 5$mT) and image the flake (Fig.~\ref{SIFig: Temp series 2}, left panel). 
At this field and temperature the odd layer (1,3) regions were FM coupled and the even-layer (2) regions were AFM coupled, resulting in zero stray magnetic field being observed. 
This configuration was maintained up at $T = 130$~K without evidence of deviation.
However at $T=138$K we observed a dramatic shift, where magnetic features appear in the previously pristine mono and tri-layer regions. 
Increasing the temperature further to $T=140$~K removed many of these features, illustrating the very small temperature window ($\delta T_{KT^*} \sim 10~K$) that this effect is observable.  

In this measurement series, one can observe a gradient in the magnetic field strength across the sample, particularly at higher temperatures (Fig.~\ref{SIFig: Temp series 2}, right panels). 
We attribute this to a temperature gradient that is induced across the sample due to joule heating from the MW antenna, which is used to drive the NV spin and is located to the right of these images, approximately 80$\mu$m away. 
Additionally, during this measurement series, the temperature readout to the main sensor was lost. 
As such, the temperature quoted was measured further away than the rest of the measurement series. 
We expect that this will lead to a larger difference in the real temperature versus the measured temperature, which could be greater than 5 degrees but is unknown.

\begin{figure*}
	\centering
	\includegraphics{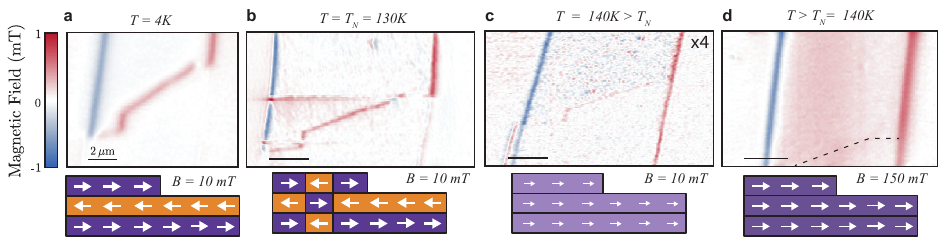}
	\caption{Temperature series for tri- and bi-layer flake. 
		\textbf{a} Magnetic image at $T = 120K$ with trilayer in the FM state and the bilayer in the AFM state. The rough state of the material is illustrated below the image.
		\textbf{b} Magnetic image at $T = 130 K \approx T_n$ showing the same state as in \textbf{a} except for one region that had the opposite sign. This was concluded to be a short term flip of the entire flake rather than a local domain, see text for details. 
		\textbf{c} Magnetic image at $T=140 K$ where both regions are in the FM state but now have zero remnant field, resulting in a reduced field strength.
		\textbf{d} Magnetic image at $T = 140 K$ measured at a higher field $B = 150 mT$ which magnetises the layers to a similar strength to before the N\'eel temperature.  
		All measurements are performed with a magnetic field pointing approximately along the CrSBr b-axis with a small $\sim 30^\circ$ out of plane component.
	}
	\label{SIFig: tri temp}
\end{figure*}

\textbf{Anisotropy transition in non-encapsulated material}
We also performed a smaller temperature scan on Sample \Sone, over a region containing both a bi- and tri-layer. At $T = 4K$ and low applied magnetic fields, the bilayer was observed to be perfectly AFM, with no observable stray magnetic field, while the tri-layer was uniformly magnetised over regions of 10s of microns, shown in Fig.~\ref{SIFig: tri temp}\textbf{a}. When heated to near the N\'eel Temperature ($T = 130 K \approx T_N$) the trilayer showed evidence of spontaneous reordering of the magnetisation direction. This is observed in the small red section of the left hand side in Fig.~\ref{SIFig: tri temp}\textbf{b}. The change in magnetic field direction indicates that for a few lines (image taken by scanning with a faster horizontal axis and a slow vertical axis) the magnetisation direction had flipped. We also observed for a previous smaller scan at the same temperature that the magnetisation direction was opposite to the scan shown in \textbf{b}. We suggest that the system was close enough to the N\'eel temperature that small local perturbations in temperature may have resulted in a flipping of the magnetisation direction through heating above $T_N$ and then cooling back down.

By increasing the temperature further to above the N\'eel Temperature ($T = 140$~K), we observed the bi-layer to become weakly FM, which results in a small stray field even for weak applied magnetic fields. 
While the tri-layer became significantly weaker in strength and formed small magnetic textures similar to the encapsulated sample (Fig.~\ref{SIFig: tri temp}\textbf{c}).

\subsection{Magnetic field dependence of the anisotropy transition}\label{Sec: SHO B}

Near $T_N$, the anisotropy transition can be suppressed (i.e. the triaxial FM phase can be stabilized) with a modest applied magnetic field. In the case of the trilayer flake shown in  Fig.~\ref{SIFig: tri temp} increasing the field to  $B = 150$ mT was enough to completely remove the magnetic textures. Unfortunately, this sample was damaged before more extensive studies of these magnetic textures could be performed. 

\begin{figure*}
	\centering
	\includegraphics{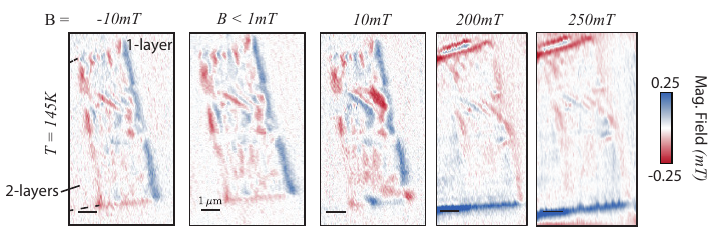}
	\caption{Magnetic field sweep at $T_c$. Series of measurements of the monolayer section of sample \Sfour around the critical temperature where long-range order is disrupted.
	}
	\label{SIFig: temp B sweep}
\end{figure*}

We did however take a series of different measurements on Sample \Sfour, shown in Fig.~\ref{SIFig: temp B sweep}, taken at $T= 145 K$. These measurements also confirm that magnetic field strength of $B>200 mT$ are sufficient to orientate the spins along the Zeeman direction. However, we also observe differences in the stray magnetic fields for small changes in the external magnetic field ($\Delta B < 10 mT$)

\section{Estimation of in-plane anisotropy}

\begin{figure*}
	\centering
	\includegraphics{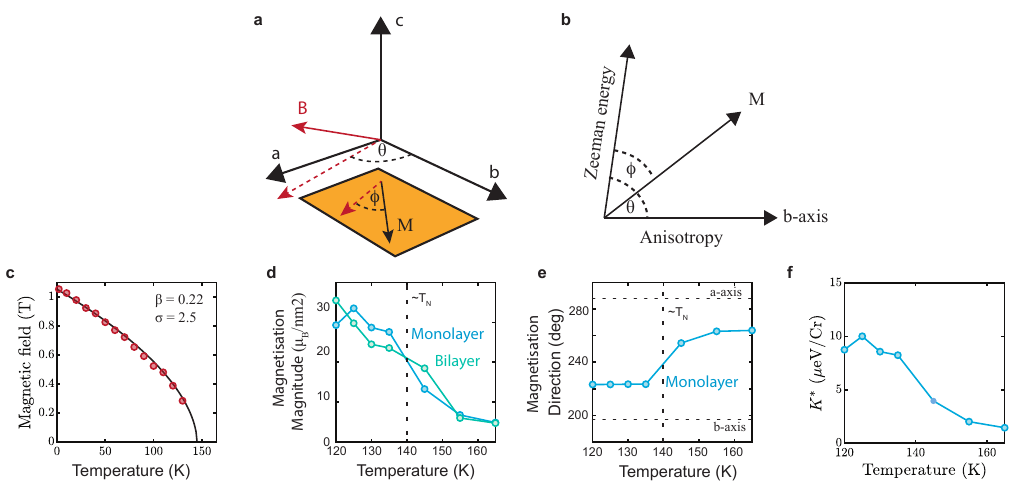}
	\caption{Method for extracting anisotropy from NV measurements. 
	\textbf{a} Illustration of a magnetic flake with an easy axis along the b-axis and a magnetic field at an arbitrary angle (red arrow, dashed for projection onto the a-b plane).  
	\textbf{b} Illustration of the energy landscape for the the magnetisation, where the magnetisation points along a direction that is given by the combination of Zeeman and anisotropy energy.  
	\textbf{c} Bulk measurement of the flipping field along the a-direction and extracted critical behaviour. 
	\textbf{d,e} Magnetisation strength (\textbf{d}) and direction (\textbf{e}) dependence as a function of temperature extracted from the measurement series discussed in Sec.~\ref{Sec: reconstruction}.
	\textbf{f} Calculated in-plane anisotropy as a function of temperature (blue) and the theoretical prediction from the Stoner-Wohlfarth model (black). 
	} 
	\label{SIFig: Anisotropy}
\end{figure*}

In order to make an estimate of the in-plane anisotropy using the temperature series we use the Stoner-Wohlfarth model~\cite{Stoner1948}, as outlined in the following. 
We model the energy of a magnetic system under the influence of a magnetic field using the expression
\begin{equation}\label{Eq: SW}
	E = K V \sin^2 (\phi - \theta) - \mu_0 M_s V B_{ext} \cos \phi,
\end{equation}
where K is the magnetocrystalline anisotropy, V is the magnet volume, $M_s$ is the saturation magnetisation, $B_{ext}$ is the external magnetic field, $\phi$ is the angle between the magnetisation direction and the external field, and $\theta$ is the angle between the applied field and the easy axis of the material. 
An illustration this situation is shown in Fig.~\ref{SIFig: Anisotropy}\textbf{a} and an energy diagram in Fig.~\ref{SIFig: Anisotropy}\textbf{b}.

To analyze our data on monolayer CrSBr, we extract the anisotropy from the angle of the magnetic field and the measured magnetization direction, using the derivative of Eq.~\ref{Eq: SW} with respect to the angle $\phi$, 
\begin{equation}
	\frac{\partial E}{\partial \phi} = 2 K V \sin (\phi - \theta)\cos (\phi - \theta) + M_s V B_{ext} \sin \phi.
\end{equation}
Then we evaluate when the derivative is zero and rearrange to get,
\begin{equation}
	K = \frac{M B_{ext} \sin\phi}{2 \sin(\phi - \theta)\cos(\phi - \theta)}.
\end{equation}
Using this equation we determine the anisotropy of the monolayer using the measured magnetisation $M$  (Fig.~\ref{SIFig: Anisotropy}\textbf{d}) and direction $\phi$,  (Fig.~\ref{SIFig: Anisotropy}\textbf{e}).
The resulting single.spin anisotropy energy as a function of temperature is shown as the blue points in Fig.~\ref{SIFig: Anisotropy}\textbf{f} and shows good qualitative agreement with the results inferred from our bulk measurements (black line).

\section{Reconstruction of magnetisation}\label{Sec: reconstruction}

\begin{figure}
	\centering
	\includegraphics{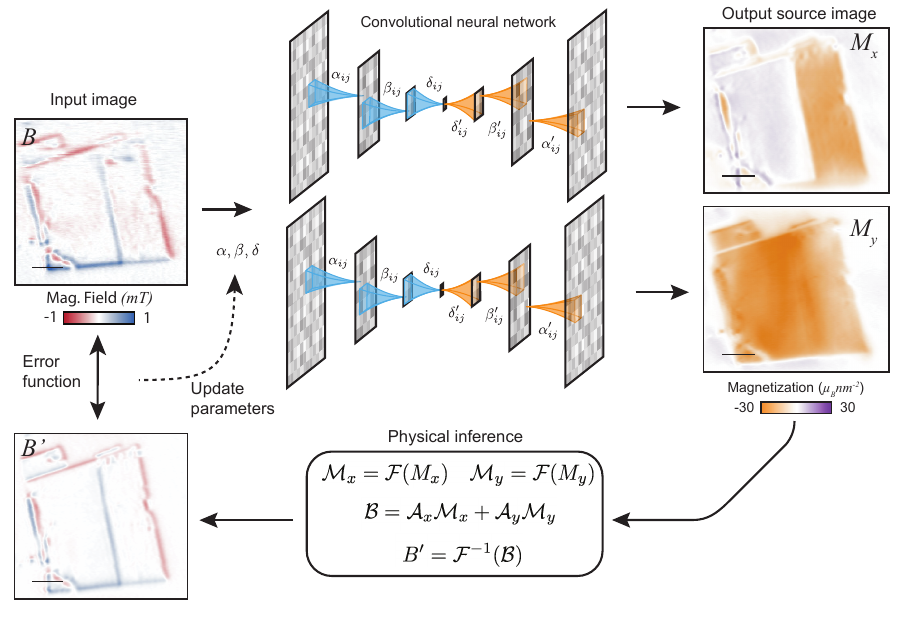}
	\caption{Illustration of the neural network reconstruction of non-uniform direction magnetisation. The input magnetic field $B$ is fed into two neural networks to produce both a $M_x$ and $M_y$ image, which is fed back to reconstruct a magnetic field image to produce an error function. The scale bar is 2 $
	mu$m}
	\label{SIFig: recon}
\end{figure}

\begin{figure*}
    \centering
    \includegraphics{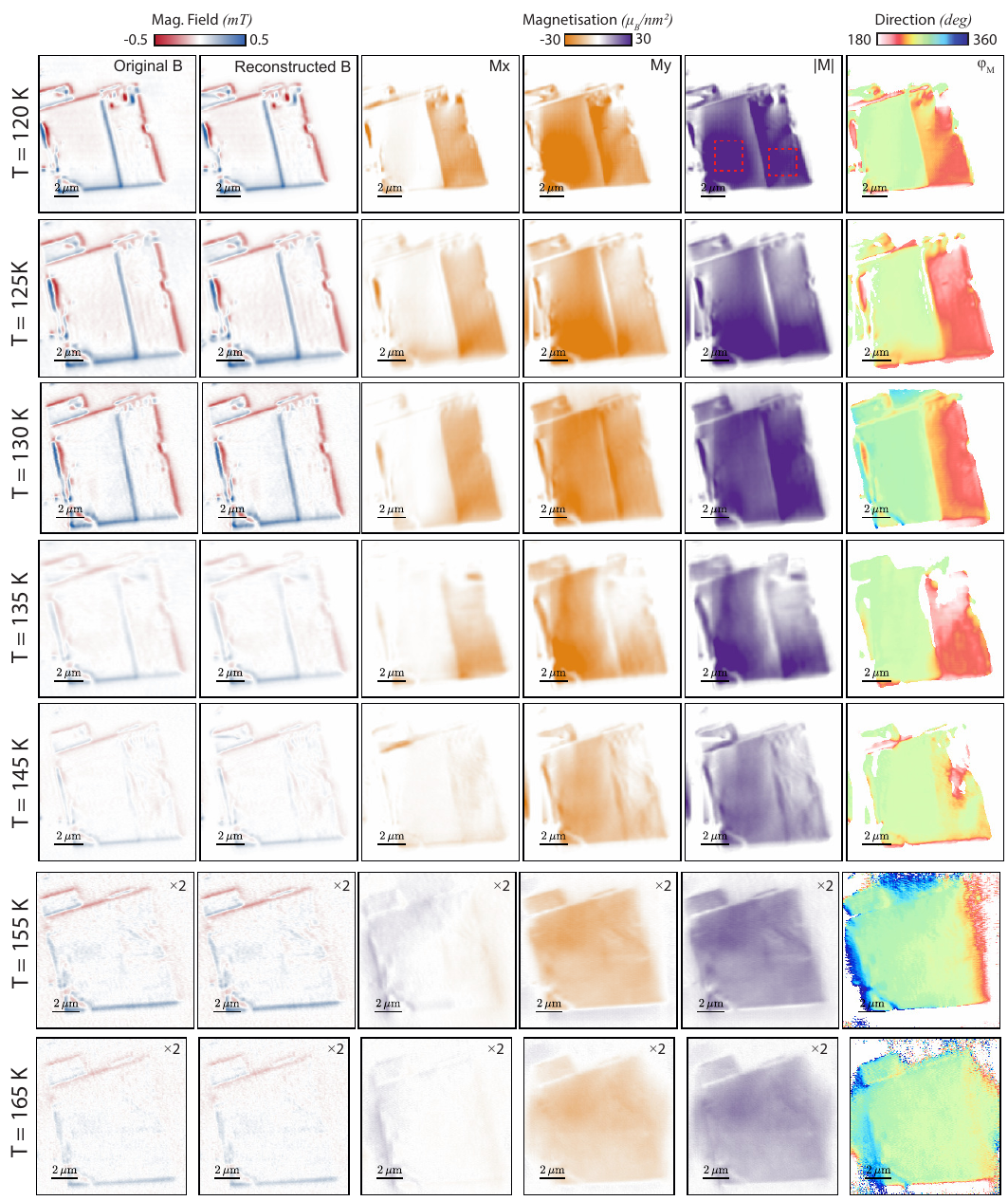}
    \caption{Magnetisation reconstruction of all the magnetic images taken at different temperatures in the main text. The rough area for the average magnetisation magnitude and direction is shown as black dashed squares on the top row. The panels from left to right are: The original magnetic field, the reconstructed magnetic field by the neural network, the magnetisation along the X-direction of the image, the magnetisation along the Y-direction of the image, the magnitude of the magnetisation, and the direction of the magnetisation. The bottom two rows have the data multiplied by two to make the data visible. The magnetisation direction has had a threshold-based background subtraction, see text for details.}
    \label{SIFig: recon data}
\end{figure*}

The reconstruction of magnetisation from a magnetic field can be performed using a Fourier space transformation where the transformation from $\mathcal{B}$ to $\mathcal{M}$ is given by a transformation matrix $\mathcal{A}$ such that, 
\begin{equation}
	\mathcal{B} = \mathcal{A} \mathcal{M}
\end{equation}
where $A$ is non-invertible but is often approximated to retrieve a reliable reconstruction~\cite{Broadway2020a}. In the case of in-plane magnetisation, such as in CrSBr, this reconstruction is difficult to perform as noise in the data often results in artifacts that are difficult to remove and an inconsistent estimation of magnetisation strength. Likewise, in traditional reconstruction the magnetisation direction is set in the reconstruction process, meaning that the magnetisation can only vary along one direction. This has two problems, first, the magnetisation direction is often not known and thus approximations of the direction can lead to further inconsistencies, and second, the magnetisation direction may vary across the image. 

To overcome these constraints we recently developed a machine learning approach that estimates the inversion transformation matrix ($\mathcal{A}^{-1}$) to produce a magnetisation image~\cite{Dubois2022}. This image is then transformed back into a magnetic field using the well-defined matrix $\mathcal{A}$ which is then compared with the measured magnetic field as an error function. Through this process, the magnetic image is fitted using the machine learning neural network, where the end result must be a valid solution to the measured magnetic field. That is, the error is minimised between the measured magnetic field and the projected magnetic field from the reconstructed magnetisation. Additionally, it allows for a reliable estimation of the uniform magnetisation direction of the material for arbitrary direction. 

This previous work was still limited by the requirement that the magnetisation direction needed to be uniform across the image. In the work in the main text, we were able to initialise the material such that the bi-layer, which had a low anisotropy, was orientated along the magnetic field direction, while the monolayer, which had a high anisotropy, was pointed along its preferred direction. In order to reconstruct this non-uniform magnetisation direction we modified the neural network from previous work to have two channels. That is, rather than generating a magnetisation image with a given magnetisation direction $\theta_M$, it produces two images one for $M_x$ and one for $M_y$. These images can have their respective magnetic fields calculated in the same manner as previously and then the magnetic fields are combined to compare with the measured magnetic image. A Schematic of the neural network is shown in Fig.~\ref{SIFig: recon}.

The reconstructed magnetisation can be used to extract the total magnetisation and the magnetisation direction for each pixel. Where we define the magnetitude as
\begin{equation}
    \abs{M} = \sqrt{M_x^2 + M_y^2}
\end{equation}
and the direction at 
\begin{equation}
    \phi_M = \arctan \left( \frac{M_y}{M_x}\right).
\end{equation}
The full reconstruction of all the temperature data is shown in Fig.~\ref{SIFig: recon data}.
To minimise the background noise two techniques are used. 
For the lower temperature case, a mask is introduced to restrict the reconstruction to produce magnetisation only inside of the flake. 
The mask is introduced by edge detection of the magnetic field. 
In the case of the higher temperature data (T = 155 and 165 K), the signal is weaker so instead in the magnetisation direction the images have a threshold applied such that 
\begin{equation}
    \phi_M = \text{NaN if } \abs{M} < 5, 
\end{equation}
which results in a white background. 

We note that the image at $T=135K$ shows a decreasing strength of the magnetic field going from the bottom of the image to the top. This is due to the diamond tip picking up dirt which increased the stand-off, and as such, prohibited reliable reconstruction of the magnetisation in this region.

\section{Bulk measurements}

To elucidate the observations made using the scanning NV magnetometer, we use a series of measurements on bulk CrSBr crystals. All bulk magnetic measurements were performed on a Quantum Design Dynacool Physical Property Measurement System (PPMS). For oriented magnetic measurements, two single crystals of CrSBr were mounted on a quartz paddle using GE varnish with the a-, b-, or c-axis aligned with the direction of the instrument's magnetic field. For subsequent orientations, the varnish was removed using a 1:1 mixture of toluene and ethanol, and the same crystals were re-mounted in a different orientation. DC and AC magnetic measurements were performed with the PPMS vibrating sample magnetometry and AC magnetometry modules, respectively.

\subsection{Bulk measurements of anisotropy}

Field-dependent magnetization measurements were collected for CrSBr along the a, b, and c axes at fixed temperatures between 2 to 130 K and in the magnetic field range 0 to 9 T. For a- (intermediate) and c- (hard) axis measurements, linear fits were applied to the low-field and high-field magnetization data.
The intercept of these two linear fits was taken as the saturation field, and the y-intercept of the high-field linear fit was taken as the saturation magnetization ($M_{sat}$). 
For b-axis measurements, the low-field fit was replaced by a fit to the linear region at the metamagnetic transition field, and the saturation field and magnetization were determined in the same way as for the a and c axes. 
The temperature-dependent anisotropy field ($H_{ani})$ for the a (c) axis was then determined as the difference between the a- (c-) axis saturation field and the b-axis saturation field. 
The effective magnetic anisotropy energy ($K^*$) for the a and c axes was then calculated at each temperature using the Stoner-Wohlfarth model~\cite{Stoner1948, Zhdanova2011}: 
\begin{equation}
	K^* = \frac{\mu_0 H_{ani} M_{sat}}{2} 
\end{equation} 

\subsection{High-Temperature AC Magnetic Susceptibility: Phase Transition Analysis}

High-temperature AC magnetic susceptibility measurements were performed to probe magnetic phase transitions near the ordering temperature. These AC magnetic susceptibility measurements were collected with the AC and DC magnetic fields oriented along the crystallographic a-axis. 
Measurements were collected for oscillating fields between 4 and 15 Oe; the AC susceptibility was found to be field-independent in this range, and all measurements shown here were collected with a 10 Oe oscillating field.  

As described in the main text, $\chi_{ac}$ has no imaginary component for the full temperature range studied here (i.e. $\chi_{ac}''$ = 0). This differs from other layered A-type antiferromagnets, such as $\text{CrCl}_{\text{3}}$~\cite{Liu2020}, for which AC susceptibility measurements reveal a high-temperature transition corresponding to short-range intralyer order, with features in both $\chi_{ac}'$ and $\chi_{ac}''$, and a lower-temperature transition corresponding to interlayer antiferromagnetic order, which only generates a feature in $\chi_{ac}'$. 
Despite evidence for ferromagnetic correlations in CrSBr as high as 200 K, the AC susceptibility data presented here suggests the absence of ferromagnetic domains above $T_N$.  

The DC-field dependent measurements shown in Figure 4d were collected with an oscillation frequency of 1000 Hz. 
For DC fields below 2000 Oe, the AC magnetic susceptibility is unchanged as a function of the DC field. For DC fields between 2000 and 3000 Oe, a high-temperature shoulder is observed, but is not well separated from the peak at $T_N$. 
As such, only data for DC fields above 3000 Oe was considered in our analysis. 
Data collected for fields above 3000 Oe were then fit using a previously established procedure~\cite{Zhao1999} to extract the critical exponents reported in the text. The dashed line in Figure 4d corresponds to the equation:
\begin{equation}
	\chi_m = e^a \left( \frac{1-T}{T_m} \right)^{-\gamma}  
\end{equation} 
with $a$ = -3.25(5), $T_m$ = 118.6(9), and $\gamma$ = 2.25(5). Here, $\chi_m$ is the susceptibility value of the field-induced maximum,  $a$ is a constant determined by the value of the order parameter $\beta$ = 0.22(2), $T_m$ is the temperature of the field-induced maximum, and $\gamma$ is the susceptibility exponent. 
Notably, the value of $T_m$ determined from these fits implies that the field-dependent features observed in AC susceptibility correspond to a zero-field ordering temperature lower than $T_N$; in other words, critical fluctuations above $T_N$ suggest a lower magnetic ordering temperature than is experimentally observed.
While further experiments are needed to better understand the origin of this behaviour, we speculate that the crossover from easy-plane to tricritical anisotropy may play some role in facilitating order in CrSBr, which may help to explain the inhomogeneous magnetization observed in the vicinity of  $T_N$. 
We further note that this discrepancy cannot be explained purely on the basis of interlayer effects, as the magnetic ordering temperature of monolayer CrSBr is largely unchanged from that of the bulk.

\subsection{Low Temperature AC Magnetic Susceptibility: Bulk Phase Wall Dynamics}

\begin{figure*}
	\centering
	\includegraphics{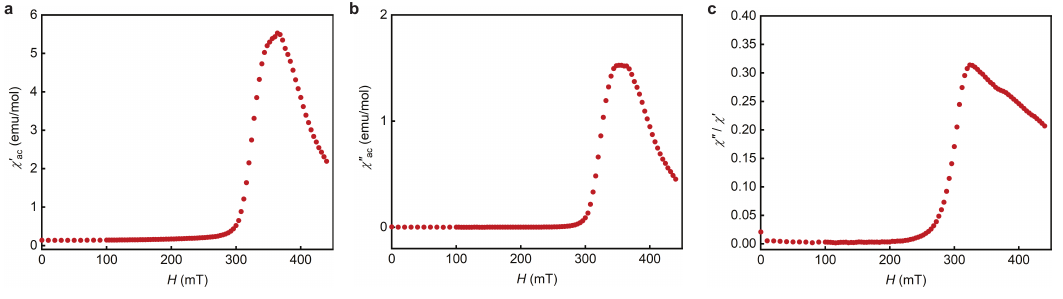}
	\caption{AC magnetic susceptibility measurements as a function of an external magnetic field.
	The in-phase (\textbf{a}), out-of-phase (\textbf{b}) and the ratio (\textbf{c}) measured at $ T = 2$K. }
	\label{SIFig: DWM chi}
\end{figure*}

Low-temperature AC magnetic susceptibility measurements were performed to probe domain wall dynamics in bulk crystals for the field-induced AFM-to-FM transition. 
Measurements were performed at 2 K with the AC and DC magnetic fields oriented along the crystallographic b-axis, with an oscillating field of 10 Oe and a drive frequency of 1000 Hz. 
At the bulk metamagnetic transition field (~0.3 T), we observe a sharp increase in $\chi_{ac}'$ and the emergence of a non-zero $\chi_{ac}''$ (Fig.~\ref{SIFig: DWM chi}).
We note that the magnetic field required to induce this transition in bulk crystals is approximately twice as large as that observed for the bilayer, consistent with the presence of two interlayer exchange interactions (instead of the single exchange interaction in the bilayer case).
The appearance of a signal in $\chi_{ac}''$ at the phase transition indicates irreversible domain wall propagation~\cite{Balnda2013},  
Given that the NV measurements clearly demonstrate that this phase transition is driven by phase-wall propagation through the material, this measurement supports the observation that the phase walls themselves are stable for a given external field. 

In the bulk crystal AC susceptibility, we additionally observe an extended tail to high fields ($>440$\,mT) in both $\chi_{ac}'$ and $\chi_{ac}''$.
This tail is particularly prominent in a plot of $\chi_{ac}''/\chi_{ac}'$, where a small secondary feature is observed near $\sim 380$\,mT. 
This secondary feature could suggest additional complexity in the spin-flip transition, such as a mix between a spin-flip and spin-flop transition, but we see no evidence for a higher-field feature in the bilayer flakes studied here.


\section{Micromagnetic simulations}
\begin{figure*}
	\centering
	\includegraphics{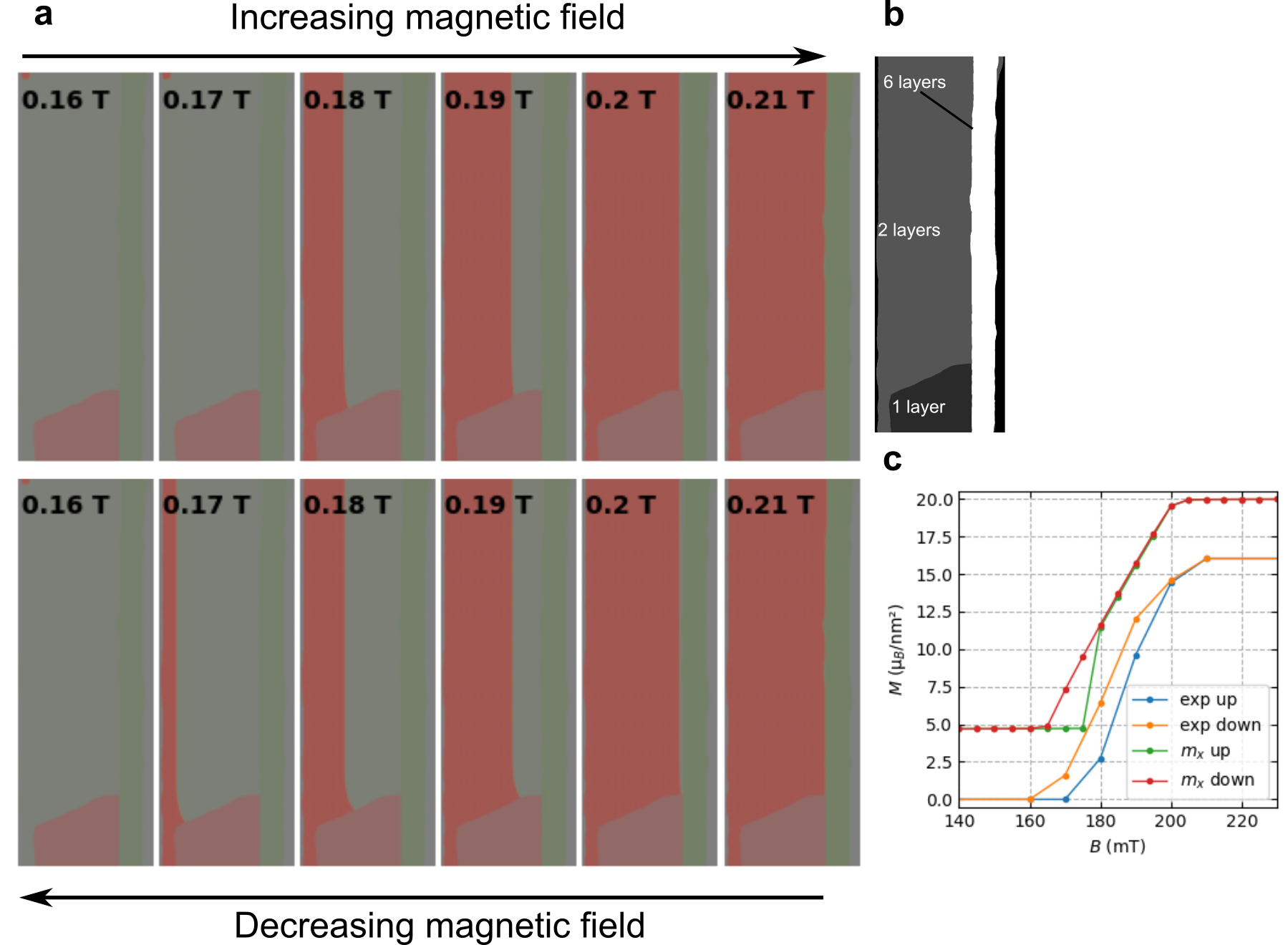}
	\caption{Simulation of domain wall hysteresis. \textbf{a} Visualisation of magnetic states of the scanned area on the CrSBr flake. Series of the domain wall movement through the material for increasing (top) and decreasing (bottom) external magnetic field. \textbf{b} Illustration of the layer geometry \textbf{c} Hysteresis of the simulated $x$ component of the total magnetic moment (offset for better visibility) and re-scaled experimental data}
	\label{SIFig: mumag}
\end{figure*}

We employ computational micromagnetic simulations~\cite{Vansteenkiste2014, Leliaert2017, Exl2014} to investigate the formation of AFM-FM phase boundaris. To do so, we simulate
the fraction of the CrSBr flake that has been imaged using the scanning NV magnetometer and is shown in Fig. 2 of the main text.  This implies the necessity to simulate lateral dimensions of a few micrometer.
In principle, atomistic simulations would be more adequate to account for the layered structure of CrSBr, but computational limits do not allow to simulate such large structures.
In our micromagnetic simulations, we mimic the layered structure of CrSBr making use of the finite difference mesh by setting the cell thickness to the thickness of one CrSBr layer. Exchange coupling between the layers is then set to be anti-ferromagnetic, while intra-layer exchange coupling is ferromagnetic.\\
To estimate material parameters, we consider a 200 nm x 200 nm x 200 nm simulation volume with periodic boundary conditions in all three dimensions, representing a bulk CrSBr crystal. The corresponding cell sizes are \SI{4}{\nano\meter} x \SI{4}{\nano\meter} x \SI{0.8}{\nano\meter}, respectively. As a starting point, the inplane exchange stiffness is estimated by $A_{ex,xy} \approx k_B T_c/(2(a+b)/2) =  \SI{2.35}{\pico\joule/\meter}$~\cite{coey_2010} using $a = \SI{0.35043}{\nano\meter}$, $b = \SI{0.47379}{\nano\meter}$, and $T_c =\SI{140}{\kelvin}$. The saturation magnetisation is set to its bulk value of $M_{sat}\approx 36~\mu_\text{B}/\si{\nano\meter\squared}$. Then, interlayer exchange coupling and strength of crystalline anisotropy axes ($a$ and $b$ axes of the material) are estimated in an iterative process: The relative strength of the two uniaxial crystalline easy anisotropy axes is taken from Fig.~3~\textbf{d} of the main text, and interlayer exchange coupling $A_{ex,z}$ is known to be negative and small compared to $A_{ex,xy}$. With this input these three parameters are adjusted until the bulk values for the saturation fields along the main crystal axes and the spin-flip field~\cite{Boix2022, Telford2022, Lopez2022} are roughly matched. This leads to $K_a =  \SI{106098}{\joule/\meter\cubed}$, $K_b = \SI{40244}{\joule/\meter\cubed}$, and $A_{ex,z} = -0.008\cdot A_{ex,xy}$.\\
For the simulations of the flake we use black and white images of each layer that are deduced from optical images of the flake as geometry, as illustrated in Fig.~\ref{SIFig: mumag}~\textbf{b}. The flake is too large to be simulated as a whole, hence, we restrict ourselves to the most relevant part, which is what has been imaged experimentally. The finite element mesh is set to a size of 825 x 2525 x 6 cells with corresponding cell sizes of \SI{4}{\nano\meter} x \SI{4}{\nano\meter} x \SI{0.8}{\nano\meter}, respectively.\\
An $x$ (along $b$ axis) gradient in $A_{ex,z}$ of about \SI{1.25}{\atto\joule/\meter/\nano\meter} is introduced to the simulation, which facilitates gradual movement of the AFM/FM phase boundary through the flake with external field. This corresponds to $A_{ex,z} = -0.008\cdot A_{ex,xy}$ on the left hand side and $A_{ex,z} = -0.0105\cdot A_{ex,xy}$ right hand side of the flake.
Further, a small `defect' in the top left corner of the flake, namely a spot with positive $A_{ex,z}$,  serves as nucleation point for the FM domain.
The saturation magnetisation is set to our measured value $M_{sat}\approx 30~\mu_{\text{B}}/\si{\nano\meter\squared}$, and $A_{ex,xy} = \SI{1.75}{\pico\joule/\meter}$ is adjusted in an other iterative process to match the field range of phase boundary movement, while the gradient field of $A_{ex,z}$ is kept at constant relative value. See Fig.~\ref{SIFig: mumag}~\textbf{c} for a comparison of experimental (re-scaled data from Fig.~\ref{SIFig: Hysteresis}~\textbf{c}) and simulated hysteresis loop ($x$ component of total magnetisation of the flake). The extremely good match of phase boundary movement field range and curve slope therein between the two data sets is owed to the before-mentioned iterative process.
Note that the field range can also be matched for other values of $A_{ex,xy}$, in this case, the relative value of $A_{ex,z}$ has to be altered.

In Fig.~\ref{SIFig: mumag} we show the gradual movement of the phase boundary through the sample in the simulated data set, which corresponds to the experimental data shown in Fig.~\ref{SIFig: Hysteresis}.
Unlike in experiment, the simulated phase boundary is perfectly vertical, except if in proximity to an edge of the sample. This is owed to the perfect orientation of the gradient in $A_{ex,z}$ with $x$. We have tested diagonal gradients leading to diagonal boundary orientations, implying that the irregular boundary in the experiment may originate in an equally irregular gradient in $A_{ex,z}$.

Potentially, other material parameters vary throughout the sample in a similar manner as $A_{ex,z}$. However, incorporating such a complex behaviour in our simulations is beyond what the given basis of information realistically allows.

The width of the AFM-FM domain wall is determined by the following procedure. First, we fit $M_z$ of the linescan shown in Fig.~2~\textbf{d} (top layer) in the main text with a $\tanh$ function and extract the domain wall parameter $\Delta$ from this. Then we use the standard definition of the domain wall width $\delta_0 = \pi\Delta$, resulting in \SI{18}{\nano\meter}. 


\end{document}